\PassOptionsToPackage{natbib}{biblatex}
\documentclass[man,12pt,floatsintext,biblatex]{apa7}
\usepackage[american]{babel}
\usepackage{csquotes}
\usepackage{graphicx}
\usepackage{amsmath,amssymb,amsfonts}
\usepackage{amsthm}
\usepackage{booktabs}
\usepackage{algorithm}
\usepackage{algpseudocode}

\title{Wild Bootstrap Inference for Non-Negative Matrix Factorization with Random Effects}
\shorttitle{NMF-RE}

\author{Kenichi Satoh}
\affiliation{Faculty of Data Science, Shiga University}

\authornote{
  \addORCIDlink{Kenichi Satoh}{0000-0003-4436-9347}

  Correspondence concerning this article should be addressed to
  Kenichi Satoh, Faculty of Data Science, Shiga University,
  Banba 1-1-1, Hikone, Shiga 522-8522, Japan.
  E-mail: kenichi-satoh@biwako.shiga-u.ac.jp

  The author declares no competing interests.
}

\abstract{
Non-negative matrix factorization (NMF) is widely used for parts-based representations, yet formal inference for covariate effects is rarely available when the basis is learned under non-negativity.
We introduce non-negative matrix factorization with random effects (NMF-RE), a mean-structure latent-variable model $Y=X(\Theta A+U)+\mathcal{E}$ that combines covariate-driven scores with unit-specific deviations.
Random effects act as a working device for modeling heterogeneity and controlling complexity; we monitor their effective degrees of freedom and enforce a df-based cap to prevent near-saturated fits.
Estimation alternates closed-form ridge (BLUP-like) updates for $U$ with multiplicative non-negative updates for $X$ and $\Theta$.
For inference on $\Theta$, we condition on $(\widehat X,\widehat U)$ and obtain fast uncertainty quantification via asymptotic linearization, a one-step Newton update, and a multiplier (wild) bootstrap; this avoids repeated constrained re-optimization.
Simulations include a targeted stress test showing that, without df control, the random-effects penalty can collapse and inference for $\Theta$ becomes degenerate, whereas the df-cap prevents this failure mode.
The non-negativity constraint induces sparse, parts-based loadings---a measurement-side variable selection---while inference on $\Theta$ identifies which covariates affect which components, providing covariate-side selection.
Longitudinal, psychometric, spatial-flow, and text examples further illustrate stable, interpretable covariate-effect inference.
}


\begin{document}
\maketitle


\section{Introduction}

Non-negative matrix factorization (NMF) produces an interpretable, parts-based
representation of multivariate non-negative data and is now routine in text
analysis, image processing, topic modeling, and genomics
\citep{lee1999,lee2000,schmidt2009,cemgil2009}.
Beyond its algorithmic appeal, NMF admits a statistical interpretation as a
latent-variable model specified through a \emph{reconstruction (mean) structure}:
each observation is approximated by an additive combination of non-negative
basis profiles and latent scores.
This emphasis on mean structure contrasts with classical factor analysis and
structural equation models (SEM), which are commonly formulated through
covariance structures \citep{joreskoggoldberger1975,bollen1989}.

Mean-structure modeling has a long tradition in longitudinal and psychometric
research, where growth-curve models represent repeated measurements via
regression on a prespecified design matrix
\citep{potthoff1964,laird1982,vonesh1987,satoh2010}.

From this perspective, NMF can be viewed as a \emph{data-adaptive} analogue of
mean-structure models: the basis matrix plays the role of a design matrix, but
is \emph{learned} from the data under non-negativity constraints rather than
fixed a priori.

Bayesian and probabilistic formulations make this data-adaptive view explicit by
casting NMF as a latent-variable model with working likelihoods and priors
\citep{schmidt2009,cemgil2009}, and statistical NMF has been developed for count
data and model selection in genomics and related settings \citep{rosales2017}.

Covariate-aware Bayesian NMF extensions have been developed for specific
likelihood settings, notably count-based mutational-signature models
\citep{robinson2019,grabski2025}.
In a frequentist direction, \citet{phelan2025} recently combined separable
(convex) NMF with post-hoc regression and nonparametric bootstrap for topic
models; their two-stage approach fixes the basis and regresses topic
proportions on covariates, but does not incorporate random effects or enforce
non-negativity on the covariate-effect matrix.
Despite this growing statistical literature, comparatively little attention has
been paid to \emph{joint} frequentist inference that integrates covariate
effects, unit-level random effects, and non-negativity constraints within a
single estimation and bootstrap framework.
Our focus is on this gap: optimization-based inference with explicit complexity
diagnostics for random effects.

A natural way to incorporate covariates is to replace free scores with a
regression $B=\Theta A$, yielding the covariate-driven approximation
$Y\approx X\Theta A$ with the ``design'' matrix $X$ learned under non-negativity
\citep{satoh2023,ding2006}.
This formulation entails an implicit form of \emph{variable selection}:
the non-negativity and low-rank constraints on $X$ jointly induce sparse,
parts-based loadings, so that each component retains only those observed
variables that contribute to it---a measurement-side selection achieved without
an explicit sparsity penalty.
When effects of covariates on latent scores in $\Theta$ are subjected to formal
inference, the resulting significance tests further identify which covariates
affect which components, providing a complementary covariate-side selection.

When $\Theta$ is the parameter of scientific interest, formal inference is
essential; yet covariance-centric modeling becomes fragile when $P$ is moderate
to large.
We therefore extend covariate-driven NMF by introducing
unit-specific random effects, leading to \emph{NMF with random effects}
(NMF-RE):
\[
Y = X(\Theta A + U) + \mathcal{E},
\qquad
\boldsymbol y_n = X(\Theta \boldsymbol a_n + \boldsymbol u_n)
+ \boldsymbol\varepsilon_n.
\]
Here $X$ is a shared non-negative basis matrix, $\Theta$ captures systematic
covariate effects on latent scores, $U$ contains unit-specific deviations in the
latent score space, and $\mathcal{E}$ is an error matrix.
This parallels mixed-effects modeling \citep{laird1982,breslow1993,kenward1997},
but differs fundamentally in that the basis $X$ is unknown and estimated under
non-negativity constraints.
We therefore view NMF-RE as a mixed-effects latent-variable model defined on a
reconstruction (mean) structure rather than on a covariance structure.

Weakly regularized random effects can absorb idiosyncratic variation and
destabilize inference for $\Theta$ \citep{efron2004,tibshirani2012}.
We therefore treat $\lambda=\sigma^2/\tau^2$ as a complexity-control device and
explicitly monitor the effective degrees of freedom consumed by $U$.

Estimation uses a block-wise scheme: random effects are updated via ridge-type
BLUP-like closed forms, while $X$ and $\Theta$ are updated via non-negative
multiplicative rules.
To prevent near-saturated regimes, we impose a diagnostic cap on the effective
degrees of freedom of $U$; the cap is a numerical safeguard and is inactive
under regular operating conditions.

Statistical inference targets $\Theta$ and is conducted \emph{conditional on the
estimated regularized solution} $(\widehat X,\widehat U)$.
Adopting a post-regularization perspective \citep{van2014,javanmard2014}, we
avoid repeated constrained re-optimization by combining (i) asymptotic
linearization, (ii) a one-step Newton update, and (iii) a multiplier (wild)
bootstrap based on weighted score contributions.
This yields computationally efficient uncertainty quantification compatible with
the algorithmic structure of NMF.

Our main contributions are:
\begin{enumerate}
\item We propose NMF-RE, a mean-structure formulation of covariate-driven NMF
with unit-specific random effects under non-negativity constraints.
\item We introduce an effective-degrees-of-freedom diagnostic for the random
effects and a df-based cap that prevents near-saturated regimes.
\item We develop post-regularization inference for $\Theta$ via the above
three-step strategy, avoiding repeated constrained optimization.
\item Through simulations (including a targeted stress test of the dfU-cap) and
four empirical studies, we demonstrate stable inference across longitudinal,
psychometric, spatial-flow, and text-analysis settings.
The empirical illustrations show that the learned $\widehat X$ achieves
measurement-side variable selection---grouping observed variables into
interpretable components without a prespecified loading pattern---while
inference on $\Theta$ provides covariate-side variable selection by identifying
which covariates significantly affect which components.
\end{enumerate}

More broadly, this work advocates a mean-structure--centric approach to inference
in constraint-based latent-variable models, combining interpretable
reconstruction structures, explicit complexity control, and implicit variable
selection through non-negative latent structure.


\section{Model}\label{sec:model}

\subsection{Data structure and notation}\label{sec:notation}
Let $Y\in \mathbb{R}_{\ge 0}^{P\times N}$ be a non-negative data matrix with
$P$ observed variables (items, words, regions, ages) and $N$ units (subjects,
documents, origin--destination nodes).
Let $A\in\mathbb{R}^{K\times N}$ collect observed covariates for the $N$ units,
with $n$th column $\boldsymbol a_n\in\mathbb{R}^K$.
We write $\mathbf{1}_m$ for an $m$-vector of ones and use $Q$ for the target
latent dimension (rank).

\paragraph{Coding of covariates.}
Because we impose elementwise non-negativity on the covariate-effect matrix
$\Theta$ (Section~\ref{sec:nmf_cov}), it is convenient to code covariates so
that $A$ is entrywise non-negative (e.g., intercepts, indicators, and
non-negative rescalings).
When a covariate takes both signs, we use the standard positive/negative-part
expansion $a=a^{+}-a^{-}$ with $a^{+}=\max(a,0)$ and $a^{-}=\max(-a,0)$, and
include both $a^{+}$ and $a^{-}$ as separate non-negative covariates.
This preserves $\Theta\ge 0$ without restricting the class of linear effects
representable by $\Theta A$.

\subsection{NMF with covariates}\label{sec:nmf_cov}
Classical NMF approximates $Y$ by a low-rank non-negative product
$Y \approx X B$, where $X\in\mathbb{R}_{\ge 0}^{P\times Q}$ is a basis matrix and
$B\in\mathbb{R}_{\ge 0}^{Q\times N}$ is a score (coefficient) matrix
\citep{lee1999,lee2000,gillis2014,gillis2020}.
To incorporate observed covariates, we parameterize the scores by a regression
on $A$,
\begin{equation}\label{eq:nmf_cov}
B = \Theta A,
\end{equation}
where $\Theta\in\mathbb{R}_{\ge 0}^{Q\times K}$ represents covariate effects on
latent scores.
The resulting reconstruction (mean) structure is
\[
Y \approx X\Theta A,
\]
which is formally analogous to the mean structure of growth-curve models
\citep{potthoff1964,laird1982,vonesh1987,satoh2010}, with the key distinction
that $X$ is \emph{learned} under non-negativity constraints rather than fixed as
a prespecified design matrix.
Related tri-factor representations appear in the machine-learning literature
\citep{ding2006}, but here we emphasize regression-style interpretation with
$\Theta$ as the primary parameter of interest.

\subsection{NMF with random effects (NMF-RE)}\label{sec:nmf_re}
In many longitudinal and multilevel settings, observed covariates do not fully
explain unit-to-unit heterogeneity.
To capture unexplained deviations while preserving interpretability, we propose
NMF with random effects (NMF-RE):
\begin{equation}\label{eq:nmf_re}
Y = X(\Theta A + U) + \mathcal{E},
\end{equation}
where $X\in\mathbb{R}_{\ge 0}^{P\times Q}$ is a shared non-negative basis,
$\Theta\in\mathbb{R}_{\ge 0}^{Q\times K}$ is a non-negative covariate-effect
matrix, $U\in\mathbb{R}^{Q\times N}$ is a matrix of unit-specific random effects,
and $\mathcal{E}\in\mathbb{R}^{P\times N}$ is an error matrix.

We treat $A$ as fixed (nonrandom) throughout.
Randomness is attributed to $U$ and $\mathcal{E}$, and hence to $Y$ through
\eqref{eq:nmf_re}.
We adopt the following working moment structure:
\begin{align}
E\!\left(\mathrm{vec}(\mathcal{E})\right) &= \mathbf{0}_{PN}, \qquad
E\!\left(\mathrm{vec}(U)\right)= \mathbf{0}_{QN}, \\
\mathrm{Var}\!\left(\mathrm{vec}(\mathcal{E})\right) &= \sigma^2 I_{PN}, \qquad
\mathrm{Var}\!\left(\mathrm{vec}(U)\right)= \tau^2 I_{QN}, \\
\mathrm{Cov}\!\left(\mathrm{vec}(U), \mathrm{vec}(\mathcal{E})\right)
&= \mathbf{0}_{QN \times PN}.
\end{align}
Under these working assumptions,
\begin{equation}\label{eq:Vy_marginal}
\mathrm{Var}(\boldsymbol y_n \mid \boldsymbol a_n)
= \tau^2 X X^\top + \sigma^2 I_P,
\end{equation}
highlighting the connection to linear mixed-effects models while keeping the
focus on the reconstruction (mean) structure.

\paragraph{Unit-level interpretation.}
For unit $n$,
\begin{equation}\label{eq:unit_model}
\boldsymbol y_n = X(\Theta \boldsymbol a_n + \boldsymbol u_n)
+ \boldsymbol{\varepsilon}_n,
\end{equation}
where $\boldsymbol a_n$ is fixed, while $\boldsymbol u_n$ and
$\boldsymbol\varepsilon_n$ are random.
Thus, columns of $X$ define $Q$ shared non-negative latent profiles, the term
$\Theta\boldsymbol a_n$ describes systematic covariate modulation of latent
scores, and $\boldsymbol u_n$ captures idiosyncratic departures.
This constitutes the mean-structure perspective introduced above,
complementing covariance-based SEM
formulations \citep{joreskog1969,joreskoggoldberger1975,bollen1989}.
As special cases, NMF-RE reduces to random-coefficient growth-curve models when
$X$ is known \citep{potthoff1964,laird1982,satoh2010}, and to standard NMF when
$A$ is absent or $U$ is suppressed \citep{lee2000,schmidt2009}.

\paragraph{Remark on signs and constraints.}
We impose $X\ge0$ and $\Theta\ge0$ for interpretability, but allow $U$ to be
real-valued; ridge regularization and df-monitoring
(Sections~\ref{sec:estimation}--\ref{sec:dfU}) prevent near-saturated behavior.

\subsection{Constraints, normalization, and identifiability}\label{sec:constraints}
NMF-type models are not fully identifiable because $X(\Theta A+U)$ is invariant
under $X\leftarrow X D^{-1}$ and $(\Theta,U)\leftarrow (D\Theta,DU)$ for any
diagonal matrix $D$ with positive diagonal entries, and under permutation of
latent components.
To fix scale, we impose a normalization on $X$.
In our implementation we use the column-sum constraint
\begin{equation}\label{eq:normalize}
X\ge 0,\quad \Theta\ge 0,\quad \mathbf{1}_P^\top X = \mathbf{1}_Q^\top,
\end{equation}
and absorb the induced scaling into $\Theta$ (and into $U$ when needed).
This normalization preserves interpretability: each column of $X$ can be viewed
as a non-negative profile (e.g., a distribution over variables), while rows of
$\Theta A$ provide the component scales across units.

Beyond scale and permutation indeterminacy, NMF generally admits only partial
identifiability; nevertheless, normalization and non-negativity often yield
stable and interpretable decompositions in practice
\citep{gillis2014,gillis2023,gillis2020}.


\section{Estimation}\label{sec:estimation}

\subsection{Complexity control via the random-effects penalty}\label{sec:tau_complexity}
We treat $\tau^2$ primarily as a complexity-control parameter:
$\lambda=\sigma^2/\tau^2$ governs the permitted unit-specific deviation.
Large $\tau^2$ (small $\lambda$) allows near-saturated fits; small $\tau^2$
shrinks $U$ toward zero.
This motivates explicit monitoring of the effective degrees of freedom consumed
by $U$ (Section~\ref{sec:dfU}) and a diagnostic cap
(Section~\ref{sec:dfUcap}).

\subsection{Working objective}\label{sec:objective}
Based on the model in Section~\ref{sec:model}, we estimate $(X,\Theta,U)$ by
minimizing the penalized least-squares criterion
\begin{equation}\label{eq:objective}
\mathcal{L}(X,\Theta,U)
= \|Y - X(\Theta A + U)\|_F^2 + \lambda \|U\|_F^2 ,
\end{equation}
subject to
\[
X \ge 0,\qquad \Theta \ge 0,\qquad \mathbf{1}_P^\top X = \mathbf{1}_Q^\top .
\]
Here $\|\cdot\|_F$ is the Frobenius norm and $\lambda$ is the complexity-control
penalty introduced in Section~\ref{sec:tau_complexity}.
We use \eqref{eq:objective} as a \emph{working} objective for computation; formal
inference for $\Theta$ is developed in Section~\ref{sec:inference}.

\subsection{Block-wise estimation scheme}\label{sec:blockwise}
Joint minimization of \eqref{eq:objective} under non-negativity constraints is
computationally challenging.
We therefore adopt a block-wise scheme that alternates updates for $U$, $X$, and
$\Theta$, holding the remaining blocks fixed.
This preserves the structure of classical NMF algorithms
\citep{lee2000,fevotte2011} while incorporating random effects.

\subsubsection{$U$-step: ridge-type update}\label{sec:Ustep}
With $X$ and $\Theta$ fixed, define $B=\Theta A$.
The $U$-subproblem is
\[
\min_U \ \|Y - X(B + U)\|_F^2 + \lambda \|U\|_F^2 .
\]
This decouples across units.
For the $n$th column,
\begin{equation}\label{eq:Ustep}
\boldsymbol u_n
= (X^\top X + \lambda I_Q)^{-1} X^\top
\bigl(\boldsymbol y_n - X \boldsymbol b_n \bigr),
\end{equation}
where $\boldsymbol y_n$ and $\boldsymbol b_n$ are the $n$th columns of $Y$ and
$B$, respectively.
Equation~\eqref{eq:Ustep} is a ridge-type update analogous to BLUP-like random
effects estimators in linear mixed-effects models \citep{laird1982,breslow1993}.
After updating $U$, we center each row to enforce zero mean across units, which
improves identifiability by preventing systematic offsets from being absorbed by
$U$.

\paragraph{Computation.}
For fixed $X$ and $\lambda$, the matrix
$(X^\top X+\lambda I_Q)^{-1}X^\top$ is common to all columns, so the $U$-update
can be implemented efficiently by forming and reusing the $Q\times Q$ solve.

\subsubsection{$X$-step: multiplicative non-negative update}\label{sec:Xstep}
With $\Theta$ and $U$ fixed, define
\[
B_U = \Theta A + U .
\]
Because $U$ is real-valued, $B_U$ may contain negative entries.
To preserve the non-negativity and numerical stability of multiplicative updates,
we use the elementwise positive part
\[
B_U^{+} = \max(B_U,0)
\]
as a stabilized surrogate in the update below.
We update $X$ using a Euclidean multiplicative rule,
\begin{equation}\label{eq:Xstep}
X \leftarrow
X \odot
\bigl(
Y (B_U^{+})^\top
\bigr)
\oslash
\bigl(
X \bigl(B_U^{+}(B_U^{+})^\top\bigr)
\bigr),
\end{equation}
where $\odot$ and $\oslash$ denote Hadamard product and division, respectively.

\paragraph{Normalization.}
After the update, we renormalize columns of $X$ to satisfy
$\mathbf{1}_P^\top X=\mathbf{1}_Q^\top$.
If $X\leftarrow XD^{-1}$ is the normalization, we apply the corresponding scaling
$(\Theta,U)\leftarrow(D\Theta,DU)$ so that the fitted mean $X(\Theta A+U)$ is
unchanged (Section~\ref{sec:constraints}).

\paragraph{Remark.}
When $U\equiv 0$ (or $B_U\ge 0$), \eqref{eq:Xstep} reduces to the standard
Euclidean multiplicative update \citep{lee2000,fevotte2011}.
The positive-part stabilization is introduced solely to keep the multiplicative
scheme well-defined when $U$ is real-valued; we monitor the working objective
\eqref{eq:objective} to ensure stable descent in practice.

\subsubsection{$\Theta$-step: covariate-driven update}\label{sec:Thetastep}
With $X$ and $U$ fixed, $\Theta$ is updated by minimizing
\[
\|Y - X(\Theta A + U)\|_F^2
\quad \text{subject to } \Theta \ge 0 .
\]
Let $Y_U = Y - XU$.
Because $Y_U$ may have negative entries, we use the positive part
$Y_U^{+}=\max(Y_U,0)$ to obtain a stable multiplicative update,
\begin{equation}\label{eq:Thetastep}
\Theta \leftarrow
\Theta \odot
\bigl(
X^\top Y_U^{+} A^\top
\bigr)
\oslash
\bigl(
(X^\top X \Theta)(A A^\top)
\bigr).
\end{equation}
This is a covariate-adapted version of multiplicative updates for non-negative
tri-factorization \citep{ding2006}, stabilized for real-valued $U$.

\subsection{Effective degrees of freedom of $U$}\label{sec:dfU}
The ridge update \eqref{eq:Ustep} introduces additional model flexibility.
For fixed $X$ and $\lambda$, the fitted random-effects contribution can be
written using the ridge ``hat'' matrix
\[
H_\lambda = X(X^\top X+\lambda I_Q)^{-1}X^\top .
\]
We define the effective degrees of freedom consumed by $U$ as
\begin{equation}\label{eq:dfU_hat}
\mathrm{df}_U = N\,\mathrm{tr}(H_\lambda).
\end{equation}
Let $d_1,\dots,d_Q$ be the eigenvalues of $X^\top X$.
Then $\mathrm{tr}(H_\lambda)=\sum_{q=1}^Q d_q/(d_q+\lambda)$ and hence
\begin{equation}\label{eq:dfU}
\mathrm{df}_U
= N \sum_{q=1}^Q \frac{d_q}{d_q + \lambda},
\end{equation}
which matches the classical ridge-regression degrees-of-freedom expression and
connects directly to complexity measures in penalized estimation
\citep{efron2004,tibshirani2012}.
This representation is convenient because $\mathrm{df}_U$ is monotone decreasing
in $\lambda$ for fixed $X$, enabling simple one-dimensional control.

\subsection{Diagnostic control via a dfU-cap}\label{sec:dfUcap}
To avoid near-saturated regimes in which $U$ absorbs most idiosyncratic variation,
we impose a diagnostic cap
\[
\mathrm{df}_U(\lambda) \le \mathrm{df}_U^{\max}.
\]
Because $\mathrm{df}_U(\lambda)$ is monotone decreasing in $\lambda$ (Section~\ref{sec:dfU}), we define
\[
\lambda_{\mathrm{cap}}(X)
= \inf\{\lambda>0:\ \mathrm{df}_U(\lambda)\le \mathrm{df}_U^{\max}\}.
\]
Whenever the current $\lambda$ violates the cap, we enforce
\[
\lambda \leftarrow \max\{\lambda,\lambda_{\mathrm{cap}}(X)\},
\]
holding $X$ fixed, and use this enforced value in the subsequent $U$-update.
Because the mapping $\lambda\mapsto \mathrm{df}_U(\lambda)$ is monotone,
$\lambda_{\mathrm{cap}}(X)$ can be obtained efficiently by a one-dimensional
search (e.g., bisection using the eigenvalue form \eqref{eq:dfU}).

\paragraph{Interpretation and practical calibration.}
The dfU-cap is a numerical safeguard, not a tuning parameter for prediction.
In practice, the cap ratio $r^{\max}=\mathrm{df}_U^{\max}/(NQ)$ is calibrated by
translating an analyst-imposed lower bound on $\lambda$ into an implied
saturation level via the eigenvalues of $X_{\mathrm{fix}}^\top X_{\mathrm{fix}}$,
where $X_{\mathrm{fix}}$ is obtained from the initialization fit with $U\equiv 0$
(details in Appendix~A).
For transparent reporting, we distinguish whether the cap was ever
\emph{activated} during optimization and whether the final solution is
\emph{binding} ($\mathrm{df}_U=\mathrm{df}_U^{\max}$) or non-binding.

\subsection{Convergence, initialization, and implementation}\label{sec:convergence}
The block-wise updates are iterated until the relative change in the working
objective \eqref{eq:objective} falls below a tolerance.
A typical iteration proceeds as follows:
(i) enforce the dfU-cap at the current $X$ (Section~\ref{sec:dfUcap});
(ii) update $U$ by \eqref{eq:Ustep} and row-center;
(iii) update $X$ by \eqref{eq:Xstep} and renormalize, applying the induced scaling
to $(\Theta,U)$;
(iv) update $\Theta$ by \eqref{eq:Thetastep}.

Initialization uses a covariate-driven NMF fit (with $U\equiv 0$) to obtain
starting values for $(X,\Theta)$, after which $U$ is introduced and updated by
ridge steps.
All updates admit closed-form or multiplicative expressions and scale to
moderate and large matrices.

\paragraph{Working scale used during optimization.}
In some implementations we update a working scale to stabilize the effective
penalty level, using
$\widehat\sigma^2_{\mathrm{iter}} = \|R\|_F^2/(PN)$ with
$R = Y - X(\Theta A + U)$,
and expressing $\lambda$ through $\widehat\sigma^2_{\mathrm{iter}}$ when needed.
This iteration-scale quantity is used only for numerical stabilization;
for statistical inference we instead use a degrees-of-freedom adjusted variance
estimator (Section~\ref{sec:score}).

A reference implementation is available in the \texttt{nmfkc} package
\citep{satoh2025nmfkc}, and all scripts needed to reproduce the numerical
results in this paper are provided at
\url{https://github.com/ksatohds/nmfre-paper}.
The complete block-wise estimation algorithm, including pseudocode for all steps
and the descent safeguard, is given in Appendix~B.

\paragraph{Variable selection as an estimation outcome.}
The non-negativity and low-rank constraints on $X$ are not designed as explicit
sparsity penalties, yet the estimated $\widehat X$ typically exhibits sparse,
parts-based structure in which each component loads on only a subset of observed
variables---an implicit measurement-side variable selection that emerges from the
estimation procedure.
When combined with formal inference for $\Theta$
(Section~\ref{sec:inference}), significance testing further identifies which
covariates affect which components, providing a complementary covariate-side
selection.


\section{Inference}\label{sec:inference}

\subsection{Scope and inferential target}\label{sec:scope}
This section develops statistical inference for the covariate-effect matrix
$\Theta$ in the NMF-RE model.
Our inferential target is $\Theta$ \emph{conditional on} the estimated
regularized representation $(\widehat X,\widehat U)$ obtained in
Section~\ref{sec:estimation}, in the spirit of post-regularization inference
\citep{van2014,javanmard2014,taylor2015}.
Robustness to working-model misspecification is pursued via sandwich and multiplier
bootstrap constructions.

\subsection{Profiled score, information, and variance scale}\label{sec:score}
Let $\widehat X$ denote the estimated basis matrix from
Section~\ref{sec:estimation}, and let $\lambda$ denote the final ridge penalty
used for the random effects.
For inference on $\Theta$, we treat $(\widehat X,\lambda)$ as fixed and profile
out $U$ via ridge regression for each candidate $\Theta$:
\[
\widehat U(\Theta)
=
\arg\min_{U\in\mathbb{R}^{Q\times N}}
\ \|Y-\widehat X(\Theta A+U)\|_F^2+\lambda\|U\|_F^2.
\]
This yields the column-wise update
\[
\widehat{\boldsymbol u}_n(\Theta)
=
(\widehat X^\top\widehat X+\lambda I_Q)^{-1}\widehat X^\top
\bigl(\boldsymbol y_n-\widehat X\Theta\boldsymbol a_n\bigr),
\qquad n=1,\dots,N,
\]
which coincides with $\widehat U$ at $\Theta=\widehat\Theta$.

\paragraph{Variance scale.}
We treat $\sigma^2$ as a nuisance scale and use a degrees-of-freedom adjusted
estimator
\begin{equation}\label{eq:sigmahat_df}
\widehat\sigma^2
=
\frac{\|Y-\widehat X(\widehat\Theta A+\widehat U)\|_F^2}
{PN-\mathrm{df}_U-\mathrm{df}_\Theta},
\end{equation}
where $\mathrm{df}_U$ is the effective degrees of freedom consumed by the ridge
random-effects update under $(\widehat X,\lambda)$ (Section~\ref{sec:dfU}).
For $\Theta$, a simple working choice is $\mathrm{df}_\Theta=QK$.
Because $\Theta\ge0$ may place some coefficients on the boundary, a more
boundary-aware alternative is to use an ``active-set'' degrees of freedom
\[
\mathrm{df}_\Theta^{\mathrm{act}}
=
\#\{(q,k):\ \widehat\Theta_{qk}>\delta\},
\]
with a small tolerance $\delta>0$ (e.g., $\delta=10^{-8}$ times a scale factor).
In our experiments, the difference between $QK$ and $\mathrm{df}_\Theta^{\mathrm{act}}$
was typically small; we report which convention is used.
In all subsequent expressions, $\sigma^2$ is replaced by its plug-in estimate
$\widehat\sigma^2$.

\paragraph{Profiled residual and score contributions.}
Let
\[
\boldsymbol r_n(\Theta)
=
\boldsymbol y_n
-
\widehat X\bigl(\Theta\boldsymbol a_n+\widehat{\boldsymbol u}_n(\Theta)\bigr)
\]
denote the profiled residual for unit $n$.
Differentiating the profiled objective with respect to $\Theta$ and using the
first-order optimality of $\widehat{\boldsymbol u}_n(\Theta)$ (envelope theorem)
yields the score contribution
\begin{equation}\label{eq:score_n}
S_n(\Theta)
= -\frac{1}{\sigma^2}
\bigl(\widehat X^\top \boldsymbol r_n(\Theta)\bigr)\boldsymbol a_n^\top,
\qquad n=1,\dots,N.
\end{equation}
We use $\mathrm{vec}(\cdot)$ to denote column-stacking vectorization; inference
targets $\mathrm{vec}(\Theta)\in\mathbb{R}^{QK}$.

\paragraph{Ridge hat matrix and shrinkage representation.}
Define the ridge hat matrix in the $\widehat X$-space,
\[
H_\lambda
=
\widehat X(\widehat X^\top \widehat X+\lambda I_Q)^{-1}\widehat X^\top.
\]
Substituting the ridge update into the residual shows that
\[
\boldsymbol r_n(\Theta)
=
(I_P-H_\lambda)\bigl(\boldsymbol y_n-\widehat X\Theta\boldsymbol a_n\bigr),
\]
so profiling out $U$ shrinks variation in the $\widehat X$-space through
$(I_P-H_\lambda)$.

\paragraph{Working information.}
If one ignores the effect of profiling out $U$ when forming curvature with
respect to $\Theta$, a Fisher-type working information matrix is
\begin{equation}\label{eq:info}
\mathcal{I}_0(\Theta)
= \frac{1}{\sigma^2}
\left(
A A^\top \otimes \widehat X^\top \widehat X
\right).
\end{equation}

\paragraph{Information reduction due to profiling out $U$.}
Profiling out $U$ reduces the effective information for $\Theta$:
\begin{equation}\label{eq:info_IH}
\mathcal{I}(\Theta)
=
\frac{1}{\sigma^2}
\left(
A A^\top \otimes
\bigl(
\widehat X^\top (I_P-H_\lambda)\widehat X
\bigr)
\right),
\end{equation}
where the inner matrix simplifies to
$\lambda\,\widehat X^\top \widehat X(\widehat X^\top \widehat X+\lambda I_Q)^{-1}$.
We use \eqref{eq:info_IH} evaluated at $\widehat\Theta$.

\subsection{Sandwich covariance estimation}\label{sec:sandwich}
Because the Gaussian working model may be misspecified, we employ a sandwich-type
covariance estimator based on score contributions.
Let
\[
\widehat{\mathcal{I}} = \mathcal{I}(\widehat\Theta),
\qquad
\widehat{\mathcal{J}}
= \sum_{n=1}^N
\mathrm{vec}\!\bigl(S_n(\widehat\Theta)\bigr)
\mathrm{vec}\!\bigl(S_n(\widehat\Theta)\bigr)^\top .
\]
The sandwich covariance estimator for $\mathrm{vec}(\widehat\Theta)$ is
\begin{equation}\label{eq:sandwich}
\widehat{\mathrm{Var}}\!\bigl(\mathrm{vec}(\widehat\Theta)\bigr)
=
\widehat{\mathcal{I}}^{-1}
\widehat{\mathcal{J}}
\widehat{\mathcal{I}}^{-1}.
\end{equation}
Finite-sample corrections (e.g., HC-type scaling) and cluster-robust variants may
be applied when unit-level score contributions exhibit heteroskedasticity or
within-cluster dependence \citep{cameron2008}.

\subsection{One-step Newton approximation}\label{sec:onestep}
Direct bootstrap inference would require repeated re-optimization under
non-negativity constraints, which is computationally expensive.
Instead, we approximate bootstrap replicates by a one-step Newton update around
$\widehat\Theta$:
\begin{equation}\label{eq:onestep}
\mathrm{vec}(\widehat\Theta^{\,\ast})
=
\mathrm{vec}(\widehat\Theta)
-
\widehat{\mathcal{I}}^{-1}
S^{\ast}(\widehat\Theta),
\end{equation}
where $S^{\ast}(\widehat\Theta)$ denotes a resampled score vector.
This construction exploits the asymptotic linearity of $\widehat\Theta$ and is
standard in post-regularization inference \citep{van2014,javanmard2014}.

\paragraph{Non-negativity in one-step replicates.}
Because \eqref{eq:onestep} is an unconstrained linear update, the resulting
$\widehat\Theta^{\,\ast}$ may violate the elementwise constraint $\Theta\ge 0$,
especially for coefficients near the boundary.
In implementation, we apply the elementwise projection
\[
\widehat\Theta^{\,\ast} \leftarrow \max(\widehat\Theta^{\,\ast},0)
\]
when forming bootstrap replicates.
For coefficients estimated at (or near) zero, boundary effects may induce
non-Gaussian and truncated finite-sample behavior; in such cases, one-sided tests
and percentile-type bootstrap summaries are typically more appropriate than
two-sided Wald normal approximations (see Section~\ref{sec:simulation}).

The asymptotic justification of the one-step map---showing that \eqref{eq:onestep} is
asymptotically equivalent to the fully iterated Z-estimator in the interior regime under
standard regularity conditions---is given in Appendix~C
\citep[see also][]{huber1967,white1980,van2014,javanmard2014,taylor2015}.

\subsection{Multiplier (wild) bootstrap for score contributions}\label{sec:wild}
We generate resampled scores using a multiplier bootstrap scheme (often referred
to as a wild bootstrap for estimating-equation contributions).
Let $\xi_1,\dots,\xi_N$ be i.i.d.\ multipliers with $E(\xi_n)=0$ and
$\mathrm{Var}(\xi_n)=1$, such as Rademacher ($\pm1$), standard normal, or
exponential-centered weights ($\xi_n=\zeta_n-1$ with $\zeta_n\sim\mathrm{Exp}(1)$)
\citep{mammen1993,cameron2008}.
In all experiments reported below, we use the exponential-centered distribution.
We form the resampled score
\[
S^{\ast}(\widehat\Theta)
=
\sum_{n=1}^N
\xi_n\,\mathrm{vec}\!\bigl(S_n(\widehat\Theta)\bigr),
\]
and substitute into \eqref{eq:onestep} to obtain bootstrap replicates
$\widehat\Theta^{\,\ast}$ without re-solving the constrained optimization problem.

Bootstrap standard errors and confidence intervals for elements of $\Theta$ are
computed from the empirical distribution of $\{\widehat\Theta^{\,\ast}\}$.
We report both sandwich-based standard errors and bootstrap-based standard errors
to provide a transparent assessment of uncertainty and of the adequacy of the
one-step approximation.

\subsection{Computational notes}\label{sec:inference_comp}
The Kronecker structure $\widehat{\mathcal{I}}=\widehat\sigma^{-2}(AA^\top\otimes F)$
with $F=\widehat X^\top(I_P-H_\lambda)\widehat X\in\mathbb{R}^{Q\times Q}$ allows the
one-step update and sandwich covariance to be computed via small $K\times K$ and
$Q\times Q$ inversions without forming the full $(QK)\times(QK)$ matrix.

\subsection{Validity, scope, and practical diagnostics}\label{sec:validity}

The proposed inference targets $\Theta$ conditional on the selected regularized
representation $(\widehat X,\widehat U)$.
Under standard regularity conditions for M/Z-estimators, the sandwich covariance
estimator \eqref{eq:sandwich} provides large-sample robust standard errors even
under a misspecified Gaussian working model, and the multiplier bootstrap combined
with the one-step Newton map yields a computationally efficient resampling
approximation \citep{huber1967,white1980,cameron2008,van2014}.
Because bootstrap replicates are projected onto $\Theta\ge0$, coefficients near the
boundary may exhibit truncation, so directional tests and percentile bootstrap
summaries are recommended in boundary regimes.
A detailed discussion of the inferential scope, regularity conditions,
boundary behavior, and practical diagnostics (including recommended reporting
items) is given in Appendix~D.


\section{Simulation study}\label{sec:simulation}

\subsection{Goal and design}\label{sec:sim_design}
This section evaluates the finite-sample behavior of the proposed
\emph{post-regularization} inference for the covariate-effect matrix $\Theta$ in
NMF-RE.
The inferential target is $\Theta$ \emph{conditional on} the fitted low-rank
representation $(\widehat X,\widehat U)$.
Because (i) estimation uses non-negativity constraints and (ii) inference is
performed after a regularized fit, the main practical question is whether the
local linearization underlying the one-step Newton approximation
(Section~\ref{sec:onestep}) yields uncertainty quantification that behaves
stably in finite samples.
Accordingly, we examine (a) the agreement between sandwich-based standard errors
(SE) and one-step multiplier-bootstrap standard errors (BSE), and (b) the size
and coverage of Wald-type tests and confidence intervals, paying special
attention to boundary behavior induced by $\Theta\ge0$.

\paragraph{Orthodont-based baseline design ($P=4$, $Q=1$).}
The simulation design mirrors the Orthodont illustration in
Section~\ref{sec:orthodont}.
We fix $P=4$ time points and rank $Q=1$.
The basis vector is set to the estimated Orthodont profile
\[
X_{\text{true}}=(0.2308,\;0.2409,\;0.2566,\;0.2717)^\top,\qquad
\mathbf{1}_P^\top X_{\text{true}}=1.
\]
The covariate matrix is $A=(\mathbf{1},\texttt{male})^\top\in\mathbb{R}^{2\times N}$.
For $N=27$, the \texttt{male} indicator follows the Orthodont sample.
For $N\in\{100,200\}$, we generate \texttt{male} i.i.d.\ Bernoulli with success
probability equal to the Orthodont male proportion.

Data are generated from the NMF-RE mechanism
\begin{equation}\label{eq:sim_dgp}
Y = X_{\text{true}}(\Theta_{\text{true}}A + U) + \mathcal{E},
\end{equation}
where $U\in\mathbb{R}^{Q\times N}$ has i.i.d.\ $N(0,\tau^2)$ entries and
$\mathcal{E}\in\mathbb{R}^{P\times N}$ has i.i.d.\ entries independent of $U$.
We set $\sigma^2=\tau^2=1$ and use the corresponding working penalty
$\lambda=\sigma^2/\tau^2=1$ as the baseline regularization level.

\paragraph{Error distributions (robustness check).}
To probe sensitivity to non-Gaussian observation noise while keeping the
variance matched to the working model, we vary only $\mathcal{E}$:
\begin{itemize}
\item \textbf{Gaussian:} $\varepsilon_{pn}\sim N(0,1)$.
\item \textbf{Centered exponential:} $\varepsilon_{pn}=\xi_{pn}-1$ with
$\xi_{pn}\sim \mathrm{Exp}(1)$, so that $E(\varepsilon_{pn})=0$ and
$\mathrm{Var}(\varepsilon_{pn})=1$ but the noise is positively skewed.
\end{itemize}
Unless noted otherwise, random effects $U$ are generated from the Gaussian model above;
thus, this robustness check targets deviations from Gaussianity in the measurement error.

\paragraph{Boundary vs.\ interior regimes.}
We consider two scenarios:
\begin{itemize}
\item \textbf{Null (boundary null):} $\Theta_{\text{true}}[\,\texttt{male}\,]=0$.
Because $\Theta\ge 0$, this null lies on the boundary, producing a truncated
finite-sample distribution for $\widehat\theta_{\texttt{male}}$.
\item \textbf{Alternative (interior alternative):} $\Theta_{\text{true}}[\,\texttt{male}\,]=9.4285$,
matching the Orthodont estimate in Table~\ref{tab:orthodont_theta}.
\end{itemize}
In both scenarios, $\Theta_{\text{true}}[\,\texttt{intercept}\,]=90.502$.

\paragraph{Fitting and inference (baseline design).}
For each $N\in\{27,100,200\}$ and each error distribution, we generate $R=1000$
Monte Carlo datasets.
Each dataset is fitted by the block-wise NMF-RE algorithm in
Section~\ref{sec:estimation}, using the dfU diagnostic cap with
$\mathrm{df}_U^{\max}/(NQ)=0.21$.
We use the same convergence tolerance and objective monitoring/safeguards as in
Section~\ref{sec:estimation}, and (when applicable) retain the best solution
across multiple initializations to mitigate non-convexity.

Inference uses the reduced information \eqref{eq:info_IH},
the sandwich covariance \eqref{eq:sandwich}, and the one-step multiplier bootstrap
(Sections~\ref{sec:onestep}--\ref{sec:wild}) with $B=1000$ resamples and
projection onto $\Theta\ge0$.
For the boundary-null scenario (Null), hypothesis testing and reported $p$-values are
computed one-sided, aligned with the constraint $\theta_{\texttt{male}}\ge0$.

\paragraph{Variance scale and degrees-of-freedom convention.}
Throughout the simulation we estimate the variance scale by \eqref{eq:sigmahat_df}.
For $\mathrm{df}_\Theta$ we use $\mathrm{df}_\Theta=QK$ (working choice); results were
qualitatively unchanged when replacing this by an active-set degrees of freedom
count (Section~\ref{sec:score}).

\paragraph{Additional stress test for random-effects saturation (dfU-cap necessity).}
To explicitly verify that dfU-cap functions as a safety device against random-effects
saturation, we run an additional targeted stress experiment in a more flexible setting
($P=4$, $Q=3$, $N=100$), where the working random-effects penalty is intentionally
weakened to promote near-saturation.
Full design details and results are reported in Appendix~E.

\subsection{Evaluation metrics}\label{sec:sim_metrics}
We report Monte Carlo bias, standard deviation (SD), and RMSE for the male-effect
estimator $\widehat\theta_{\texttt{male}}$.
We also report Monte Carlo averages of SE and BSE.
For hypothesis testing, we use a Wald-type statistic based on either SE or BSE.
For the boundary null (Null), the main procedure is the one-sided test
$H_0:\theta_{\texttt{male}}=0$ vs.\ $H_1:\theta_{\texttt{male}}>0$.

For interval estimation, we report coverage of nominal 95\% Wald intervals based on
SE and BSE, and coverage of the 95\% percentile interval constructed from the
one-step multiplier-bootstrap replicates.
We monitor random-effects saturation via $\mathrm{df}_U/(NQ)$:
in the baseline $Q=1$ study we summarize the Monte Carlo mean of $\mathrm{df}_U/(NQ)$,
while in the targeted $Q=3$ stress test we additionally report an upper-tail summary,
dfU$_{0.99}$, the 99th percentile of $\mathrm{df}_U/(NQ)$ over Monte Carlo runs.

\subsection{Results}\label{sec:sim_results}

The full Monte Carlo results are reported in Table~F1 (Appendix~F).
The simulation supports the central claim of this paper:
\emph{conditional} inference for $\Theta$ after a regularized NMF-RE fit can be
made stable in a controlled low-rank setting.
Under the Alternative (interior regime), SE and BSE are close and coverages approach
nominal levels as $N$ increases, confirming that the one-step linearization is
adequate when coefficients are away from the boundary.
Under the Null (boundary regime), BSE is systematically smaller than SE due to
projection onto $\Theta\ge0$, but the one-sided rejection rates remain close to
5\% and coverages are slightly conservative.
Replacing Gaussian errors by centered exponential errors yields qualitatively
similar patterns, suggesting robustness to moderate non-Gaussianity.
A detailed interpretation is provided in Appendix~F.

The stress test results (Appendix~E) show that without dfU-cap, the fitted penalty
collapses ($\widehat\lambda\approx 0$) and inference for $\Theta$ becomes degenerate,
whereas a strict cap prevents this collapse and, under the Null, maintains
near-nominal size and coverage.
(Under the Alternative the intentionally weak initial penalty combined with the
higher-dimensional random-effects space induces substantial bias in
$\widehat\Theta$ under all cap settings; this is expected for a stress
scenario designed to probe failure modes rather than routine performance.)
In the baseline $Q=1$ design, the mean saturation ratio
$\mathrm{df}_U/(NQ)$ stays near $0.20$ under the default cap $\mathrm{df}_U^{\max}/(NQ)=0.21$,
indicating that $U$ is kept in a moderately regularized regime.


\section{Empirical illustrations}\label{sec:examples}

We illustrate NMF-RE with four empirical examples of increasing complexity:
Orthodont growth data, Holzinger--Swineford cognitive tests, origin--destination
flows (OD--HUB), and a topic model.
Table~\ref{tab:example_summary} provides a diagnostic summary of model sizes
and the random-effects operating regime.
The saturation ratio $r=\mathrm{df}_U/(NQ)$ measures random-effects flexibility
(0 = strongly regularized; 1 = near-saturated); the cap ratio
$r^{\max}=\mathrm{df}_U^{\max}/(NQ)$ is calibrated as in Section~\ref{sec:dfUcap}
(see also Appendix~A).
All reported solutions are non-binding ($\mathrm{df}_U<\mathrm{df}_U^{\max}$).

\begin{table}[t]
\centering
\caption{Summary of model sizes and random-effects magnitude across empirical examples.
Here $N$ is the number of units, $P$ the number of observed variables, $Q$ the latent dimension,
and $NQ$ the total number of random-effects entries.
We report $\mathrm{df}_U$ and the saturation ratio $r=\mathrm{df}_U/(NQ)$, together with the cap ratio
$r^{\max}=\mathrm{df}_U^{\max}/(NQ)$ used for the dfU-cap (Section~\ref{sec:dfUcap}).
The column ``Cap'' indicates whether the dfU-cap was ever triggered during optimization, and
$\lambda$ is the final (possibly enforced) ridge penalty used in the reported fit.}
\label{tab:example_summary}
\begin{tabular}{lrrrrrrrrr}
\toprule
Dataset & $N$ & $P$ & $Q$ & $NQ$ &
$\mathrm{df}_U$ & $r$ & $r^{\max}$ & $\lambda$ & Cap \\
\midrule
Orthodont   & 27  & 4  & 1 & 27  & 5.42   & 0.201 & 0.21 & 1.00   & No  \\
HS1939      & 300 & 9  & 3 & 900 & 861.93 & 0.958 & 0.96 & 0.0104 & No  \\
OD--HUB     & 47  & 43 & 4 & 188 & 2.38   & 0.013 & 0.02 & 6.75   & Yes \\
Topic model & 59  & 67 & 3 & 177 & 1.30   & 0.007 & 0.02 & 3.30   & Yes \\
\bottomrule
\end{tabular}
\end{table}

In subsequent subsections, we report inference for $\Theta$ and interpret the estimated
components in light of both the learned profiles ($X$) and the saturation diagnostics $(r, r^{\max})$.


\subsection{Orthodont growth data}\label{sec:orthodont}

The Orthodont growth data consist of pituitary--pterygomaxillary fissure
distances for $N=27$ children (16 males, 11 females) at ages 8, 10, 12, and
14 \citep{potthoff1964,laird1982,nlme}, yielding $P=4$ repeated measurements
with covariate $\boldsymbol a_n=(1,\mathrm{male}_n)^\top$.

\paragraph{Benchmark: linear mixed-effects modeling (LMM).}
As an orientation benchmark, we also fitted a standard LMM with random intercept
and slope (see Appendix~G for details).
The LMM confirms that the Orthodont data exhibit both substantial subject-level
heterogeneity and sex-dependent mean structure, including a significant sex
difference at age 11 and a sex-dependent growth rate.

\paragraph{NMF-RE specification ($Q=1$).}
With $Q=1$, the NMF-RE mean structure reduces to
$\boldsymbol y_n
=\boldsymbol x\,(\theta_0+\theta_{\mathrm{male}}\mathrm{male}_n+u_n)
+\boldsymbol\varepsilon_n$
with $\boldsymbol x\ge 0$, $\mathbf{1}_P^\top\boldsymbol x=1$;
here $\boldsymbol x$ is a non-negative trajectory shape learned from the data.

\begin{figure}[t]
\centering
\includegraphics[width=0.95\linewidth]{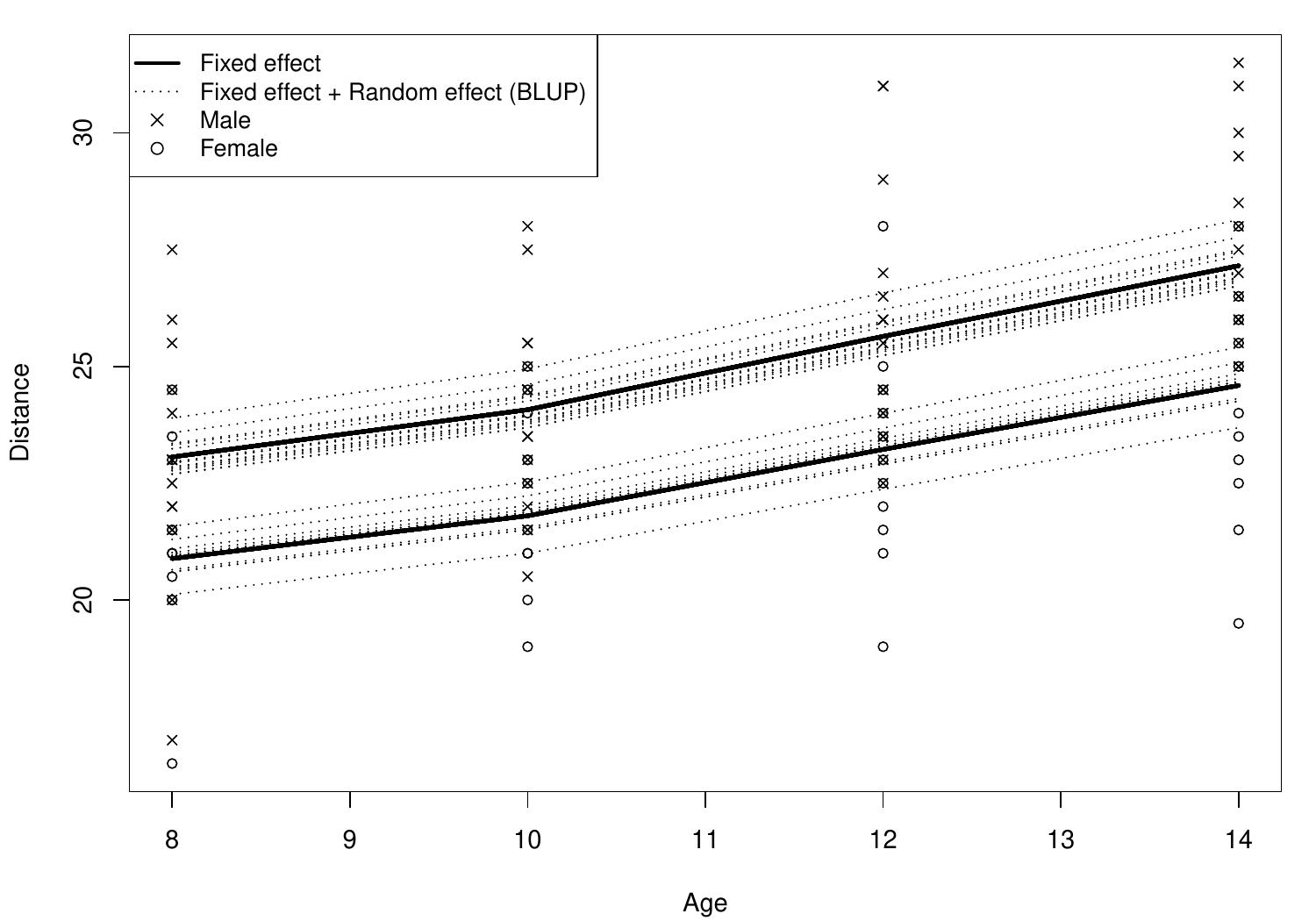}
\caption{Orthodont growth data ($Q=1$): observed measurements and fitted curves.
Solid lines show fitted values from the fixed-effects component
$\widehat X\widehat\Theta \boldsymbol a_n$, and dashed lines show BLUP-based fits
$\widehat X(\widehat\Theta \boldsymbol a_n + \widehat{\boldsymbol u}_n)$.
Points indicate observations (circles: female; crosses: male).}
\label{fig:orthodont_fit}
\end{figure}

\paragraph{Estimated basis profile and fitted curves.}
For $Q=1$, the estimated basis vector $\widehat X$ assigns non-negative weights
to ages 8, 10, 12, and 14 years:
\[
\widehat X
=
\bigl(
0.2308,\;0.2409,\;0.2566,\;0.2717
\bigr)^\top,
\qquad
\mathbf{1}_P^\top \widehat X = 1.
\]
The increasing pattern indicates that the learned trajectory places
progressively more mass on later ages, consistent with monotone growth.
Figure~\ref{fig:orthodont_fit} displays the observed measurements together with
two fitted curves for each subject: the fixed-effects component
$\widehat X\widehat\Theta \boldsymbol a_n$ and the BLUP-based fit
$\widehat X(\widehat\Theta \boldsymbol a_n + \widehat{\boldsymbol u}_n)$.
The difference between the two curves represents subject-specific deviations
captured by the random effects.

\paragraph{Random-effects regime (df diagnostic).}
For this example, the random effects operate in a regular (non-saturated) regime.
The effective degrees of freedom for the random effects is
$\mathrm{df}_U/(NQ)=0.201$, while the diagnostic cap was set to
$\mathrm{df}_U^{\max}/(NQ)=0.21$.
The cap was not activated during optimization, and hence does not constrain the
reported solution (Table~\ref{tab:example_summary}).

\paragraph{Inference for covariate effects.}
Table~\ref{tab:orthodont_theta} reports the estimated covariate effects and their
standard errors.
$p$-values are computed using a one-sided test for the boundary null
$H_0:\theta=0$ versus $H_1:\theta>0$ under the non-negativity constraint
$\Theta\ge 0$.
Both the intercept and the male effect are detected as statistically significant.
Moreover, the bootstrap-based standard errors are close to the sandwich-based
standard errors, supporting the adequacy of the one-step multiplier (wild)
bootstrap approximation in this setting.

\begin{table}[t]
\centering
\caption{Estimated covariate effects for the Orthodont growth data ($Q=1$).
Standard errors are reported from the sandwich estimator (SE) and from the
multiplier (wild) bootstrap based on one-step Newton updates (BSE).
The Wald statistic is $z=\text{Estimate}/\text{SE}$, and
$p$-values are one-sided for $H_0:\theta=0$ vs.\ $H_1:\theta>0$ under $\Theta\ge 0$.}
\label{tab:orthodont_theta}
\begin{tabular}{lrrrrr}
\toprule
Covariate & Estimate & SE & BSE & $z$ & $p$ \\
\midrule
Intercept & 90.502 & 2.471 & 2.450 & 36.62 & $<0.001$ \\
Male      &  9.428 & 3.056 & 2.975 &  3.09 & 0.0010 \\
\bottomrule
\end{tabular}
\end{table}

\paragraph{Summary.}
Unlike the LMM, which requires a prescribed time basis, NMF-RE learns the trajectory
shape from the data under non-negativity, providing a parsimonious mean-structure
summary.
This data-driven learning of $\widehat X$ constitutes measurement-side variable
selection: the non-negative basis automatically identifies which time points
contribute to the growth component.
On the covariate side, inference on $\Theta$ selects the sex effect as significant
($p=0.001$).
The random-effects component is far from saturation, with close SE--BSE agreement
(see Appendix~G for a detailed comparison with the LMM benchmark).


\subsection{Holzinger--Swineford cognitive test data}\label{sec:hs1939}

The Holzinger--Swineford (1939) data consist of $P=9$ cognitive tests for $N=300$
students \citep{holzinger1939,joreskog1969}.
The classical confirmatory factor analysis of this dataset posits a three-factor
structure (visual, textual, speed); we use NMF-RE to examine whether a comparable
structure can be \emph{learned} under non-negativity.
As covariates, we use reversed age ($\texttt{age.rev}$), sex ($\texttt{sex.2}$),
and school indicator ($\texttt{school.GW}$), all rescaled to $[0,1]$.

\paragraph{NMF-RE specification ($Q=3$).}
Motivated by the classical three-factor structure, we set $Q=3$.
The random effects operate in a near-saturated but controlled regime
($\mathrm{df}_U/(NQ)=0.958$, cap $0.96$; Table~\ref{tab:example_summary}),
signaling substantial subject-level heterogeneity beyond the observed covariates.
The cap was not activated, so the reported solution is not forced to sit on the
boundary.

\begin{figure}[t]
\centering
\includegraphics[width=0.95\linewidth]{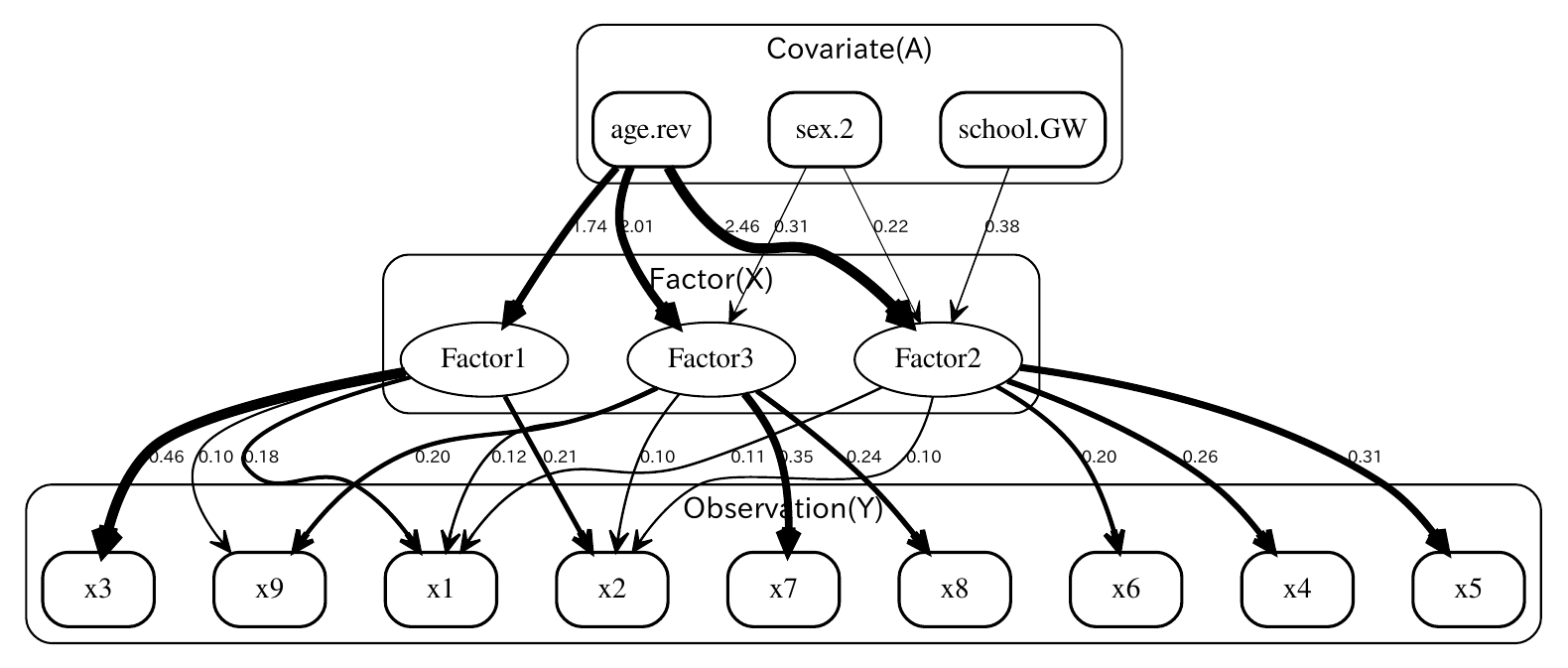}
\caption{Holzinger--Swineford cognitive test data ($Q=3$):
estimated structure linking observed tests ($Y$),
latent components ($X$), and covariates ($A$) under the NMF-RE model.}
\label{fig:hs1939_graph}
\end{figure}

\paragraph{Learned factor structure (Figure~\ref{fig:hs1939_graph}).}
The learned $\widehat X$ recovers the familiar three-domain grouping: a component
dominated by visual tests ($x_1$--$x_3$), one by textual tests ($x_4$--$x_6$),
and one by speed tests ($x_7$--$x_9$), without a prespecified loading pattern
(Figure~\ref{fig:hs1939_graph}).
Because $\widehat X$ is entrywise non-negative and column-normalized, its entries
admit a direct parts-based interpretation as relative test contributions.

\paragraph{Inference for covariate effects (Table~\ref{tab:hs1939_theta}).}
Age effects are strongly significant across all three components.
Sex and school effects are component-specific: $\texttt{sex.2}$ is detected for
Factors~2 and~3 but not for Factor~1, while $\texttt{school.GW}$ primarily
affects Factor~2.
SE--BSE agreement holds even under the high saturation ratio.

\begin{table}[t]
\centering
\caption{Estimated covariate effects for the Holzinger--Swineford cognitive test
data ($Q=3$) under NMF-RE. Format as in Table~\ref{tab:orthodont_theta}.}
\label{tab:hs1939_theta}
\begin{tabular}{llrrrrr}
\toprule
Covariate & Basis & Estimate & SE & BSE & $z$ & $p$ \\
\midrule
age.rev   & Factor1 & 1.738 & 0.089 & 0.087 & 19.54 & $<0.001$ \\
age.rev   & Factor2 & 2.456 & 0.100 & 0.100 & 24.46 & $<0.001$ \\
age.rev   & Factor3 & 2.015 & 0.078 & 0.077 & 25.74 & $<0.001$ \\
sex.2     & Factor1 & 0.000 & 0.059 & 0.036 &  0.00 & 0.5000 \\
sex.2     & Factor2 & 0.220 & 0.071 & 0.069 &  3.10 & 0.0010 \\
sex.2     & Factor3 & 0.309 & 0.060 & 0.059 &  5.14 & $<0.001$ \\
school.GW & Factor1 & 0.000 & 0.058 & 0.032 &  0.00 & 0.5000 \\
school.GW & Factor2 & 0.378 & 0.075 & 0.075 &  5.05 & $<0.001$ \\
school.GW & Factor3 & 0.000 & 0.069 & 0.042 &  0.00 & 0.5000 \\
\bottomrule
\end{tabular}
\end{table}

\paragraph{Supplementary benchmark: confirmatory MIMIC / latent regression in SEM.}
As a covariance-based SEM benchmark, we also fitted a confirmatory MIMIC model
\citep{joreskoggoldberger1975,bollen1989} using \texttt{lavaan} \citep{lavaan}
with the textbook three-factor specification and robust standard errors (see Appendix~G
for full results).
The MIMIC analysis confirms that all three covariates have nontrivial impact on the
latent structure under the confirmatory specification.
One notable difference is that the MIMIC model detects a significant \emph{negative}
sex effect on the visual factor, which the non-negativity constraint $\Theta\ge 0$
in NMF-RE projects to zero; this illustrates that $\Theta\ge 0$ trades the ability
to detect sign-unrestricted effects for interpretability and sparsity.

\paragraph{Summary.}
Unlike confirmatory SEM, NMF-RE learns the measurement structure directly from data
under non-negativity, recovering the three-domain grouping without a prespecified loading
pattern.
This illustrates measurement-side variable selection: the nine tests are partitioned
into visual, textual, and speed components by $\widehat X$ alone.
On the covariate side, inference on $\Theta$ performs component-specific covariate
selection---age is universally significant, whereas sex and school effects are
detected only for specific factors (Table~\ref{tab:hs1939_theta}).
The random-effects regime is near-saturated but non-binding, with close SE--BSE
agreement (see Appendix~G for a detailed MIMIC comparison).


\subsection{Origin--destination flow data with hub-based covariates}\label{sec:odhub}

We consider an origin--destination (OD) flow table from the Japanese National Road
Traffic Census \citep{estat2021}.
Let $F\in\mathbb{R}_{\ge 0}^{47\times 47}$ denote the OD table; we analyze the
log-transformed $\log(1+F_{ij})$.
Four metropolitan hubs (Tokyo, Aichi, Osaka, Fukuoka) are selected as covariates
($K=4$), with destinations as units ($N=47$).
The response matrix $Y\in\mathbb{R}^{43\times 47}$ collects outbound flows from the
remaining $P=43$ non-hub origins, and $A\in\mathbb{R}^{4\times 47}$ collects hub
outbound flows.

\paragraph{NMF-RE specification ($Q=4$).}
With $Q=4$ matching the number of hubs, the random effects are strongly regularized
($\mathrm{df}_U/(NQ)=0.013$; Table~\ref{tab:example_summary}), indicating that
most variation is explained by the hub-based mean structure.

\begin{figure}[t]
\centering
\includegraphics[width=0.95\linewidth]{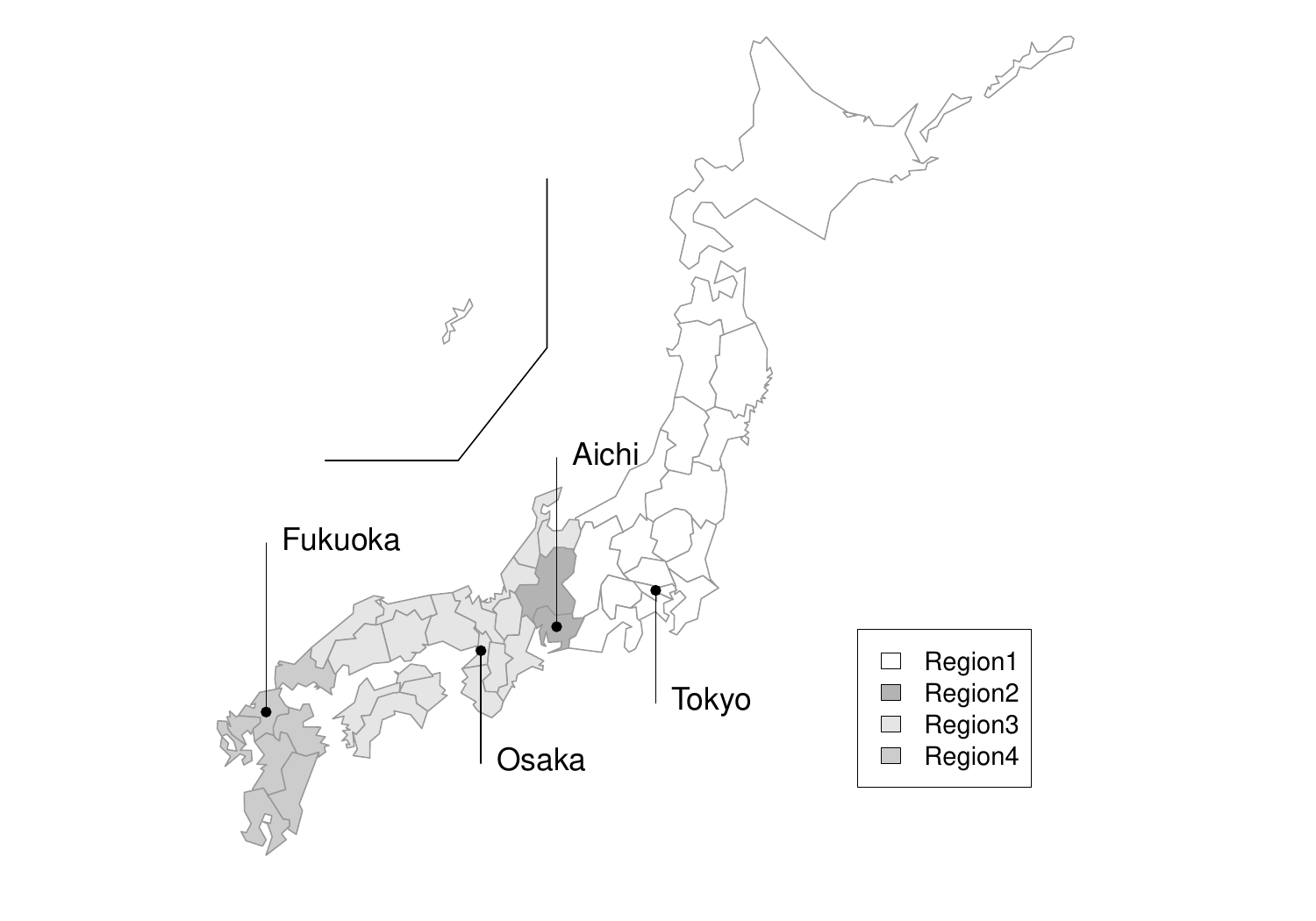}
\caption{OD--HUB data ($Q=4$): spatial visualization of the latent components identified by the
NMF-RE model. Each destination prefecture is colored according to the dominant latent component
based on the fitted latent scores.}
\label{fig:odhub_map}
\end{figure}

\paragraph{Probability-like scores and spatial visualization.}
Because fitted scores are non-negative, normalizing $\widehat B=\widehat\Theta A$
to column sums yields simplex-like representations from which a hard clustering is
obtained by $\arg\max_q$.
Figure~\ref{fig:odhub_map} colors each destination by its dominant component;
each component aligns primarily with one hub, yielding a clear geographic partition.

\paragraph{Inference for hub effects (Table~\ref{tab:odhub_theta}).}
$\widehat\Theta$ exhibits a near one-to-one hub--component association: each
component has a single dominant hub coefficient, with cross-loadings at exact zero
under $\Theta\ge0$.
SE--BSE agreement holds for the dominant coefficients; boundary coefficients show
smaller BSE due to projection onto $\Theta\ge0$.

\begin{table}[t]
\centering
\caption{Estimated covariate effects for the OD--HUB data ($Q=4$).
Format as in Table~\ref{tab:orthodont_theta}.}
\label{tab:odhub_theta}
\begin{tabular}{llrrrrr}
\toprule
Covariate & Basis & Estimate & SE & BSE & $z$ & $p$ \\
\midrule
Tokyo    & Region1 & 8.896 & 0.801 & 0.763 & 11.11 & $<0.001$ \\
Tokyo    & Region2 & 0.000 & 0.302 & 0.165 &  0.00 & 0.5000 \\
Tokyo    & Region3 & 0.000 & 0.292 & 0.163 &  0.00 & 0.5000 \\
Tokyo    & Region4 & 0.000 & 0.250 & 0.153 &  0.00 & 0.5000 \\
Aichi    & Region1 & 0.000 & 0.597 & 0.354 &  0.00 & 0.5000 \\
Aichi    & Region2 & 4.381 & 0.633 & 0.642 &  6.92 & $<0.001$ \\
Aichi    & Region3 & 0.000 & 0.524 & 0.309 &  0.00 & 0.5000 \\
Aichi    & Region4 & 0.000 & 0.503 & 0.265 &  0.00 & 0.5000 \\
Osaka    & Region1 & 0.000 & 0.546 & 0.295 &  0.00 & 0.5000 \\
Osaka    & Region2 & 0.000 & 0.322 & 0.197 &  0.00 & 0.5000 \\
Osaka    & Region3 & 6.895 & 0.637 & 0.656 & 10.82 & $<0.001$ \\
Osaka    & Region4 & 0.000 & 0.413 & 0.230 &  0.00 & 0.5000 \\
Fukuoka  & Region1 & 0.000 & 0.483 & 0.200 &  0.00 & 0.5000 \\
Fukuoka  & Region2 & 0.000 & 0.190 & 0.095 &  0.00 & 0.5000 \\
Fukuoka  & Region3 & 0.000 & 0.520 & 0.254 &  0.00 & 0.5000 \\
Fukuoka  & Region4 & 5.575 & 0.373 & 0.374 & 14.94 & $<0.001$ \\
\bottomrule
\end{tabular}
\end{table}

\paragraph{Supplementary benchmark: projection pursuit regression (PPR).}
As a flexible regression benchmark, we also fitted multivariate PPR
\citep{friedman1981,friedman1984} with $M=4$ ridge terms (see Appendix~G for details).
The estimated projection directions mix the four hubs within each term, so interpretability
hinges on ridge-function diagnostics rather than a direct hub-to-component correspondence.

\paragraph{Summary.}
NMF-RE yields a structured parts-based decomposition with non-negative scores that enable
probability-like representations and transparent geographic clustering
(Figure~\ref{fig:odhub_map}; see Appendix~G for a detailed PPR comparison).
Measurement-side selection assigns each destination to its dominant hub region
via $\widehat X$, while covariate-side selection reveals a sharp one-to-one mapping
between origin hubs and regional components
(Table~\ref{tab:odhub_theta}): each hub covariate loads exclusively on one component,
with all off-diagonal effects estimated at zero.
The very small saturation ratio implies that the partition is essentially driven by the
fixed-effects mean structure; the dfU-cap is triggered transiently but the final solution
is non-binding.


\subsection{Topic model with historical covariates}\label{sec:topic}

As a final illustration, we apply the proposed NMF-RE framework to a text analysis
problem based on U.S.\ presidential inaugural addresses.
Following standard preprocessing in \textsf{R} using \texttt{quanteda}
\citep{quanteda}, we represent the corpus as a document--term matrix of
non-negative word counts.
We retain words that appear at least 100 times across the corpus, yielding a moderate
vocabulary size ($P=67$) over $N=59$ documents.

\paragraph{Historical covariates (era indicators) and coding.}
To examine systematic historical shifts in thematic structure, we introduce three
mutually exclusive era indicators as covariates:
(i) Early (1775--1865; independence to the Civil War),
(ii) Industrial (1865--1945; industrialization and overseas expansion), and
(iii) Postwar (1945--; post--World War~II).
These indicators form the rows of $A$ \emph{without an intercept}, so that each
document belongs to exactly one era.
Under this coding, each column of $\widehat\Theta A$ equals the era-specific expected
latent-score vector, and the fixed-effects reconstruction $\widehat X\widehat\Theta A$
induces a step-function (piecewise-constant) pattern across eras when documents are
ordered chronologically.

\paragraph{NMF-RE specification and random-effects regime.}
We fit NMF-RE with latent dimension $Q=3$, interpreting each column of the fitted basis
$\widehat X$ as a topic-specific word profile.
Let $\widehat B=\widehat\Theta A$ denote the fixed-effects latent-score matrix.
For this dataset, random effects are strongly regularized:
the saturation ratio is $\mathrm{df}_U/(NQ)=0.007$ with diagnostic cap
$\mathrm{df}_U^{\max}/(NQ)=0.02$, and the cap was activated at least once during
optimization (Table~\ref{tab:example_summary}).
Thus, most systematic variation is attributed to the fixed-effects mean structure,
while remaining within-era idiosyncrasies are heavily shrunk through $\widehat U$.

\paragraph{Learned topics.}
Each column of $\widehat X$ is a topic-specific word profile; words with
$\widehat X_{pq}/\sum_{q'}\widehat X_{pq'}>0.5$ are taken as representative.
The three components are (illustratively):
Topic~1 (institutional/legal), Topic~2 (union/constitutional/foreign), and
Topic~3 (modern civic rhetoric).

\begin{figure}[t]
\centering
\includegraphics[width=0.95\linewidth]{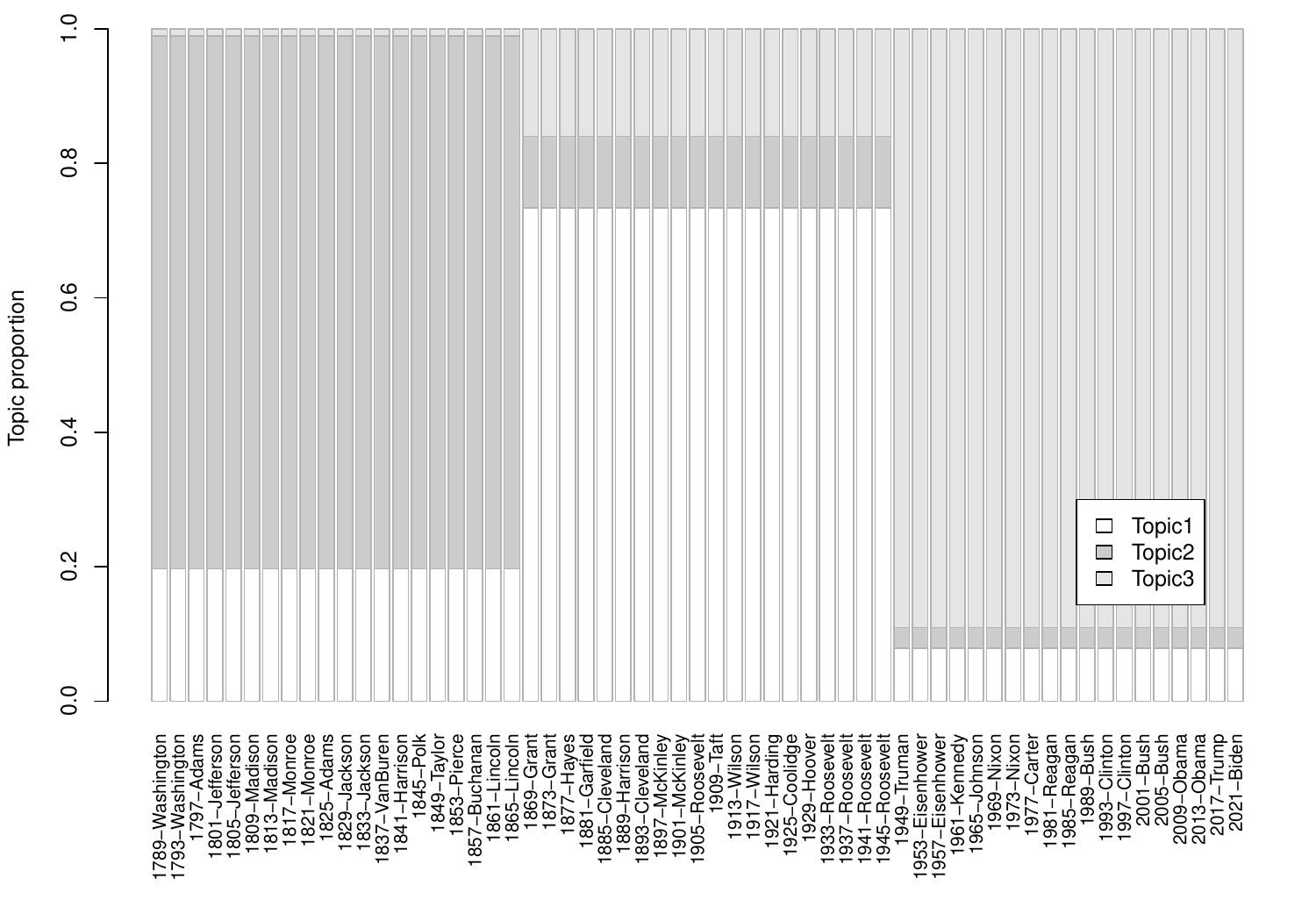}
\caption{Topic model with historical covariates ($Q=3$): estimated topic proportions based on
the fixed-effects component $\widehat X\widehat\Theta A$.
Each bar corresponds to one inaugural address (document), illustrating systematic shifts
in topic prevalence across historical eras.}
\label{fig:topic_fixed}
\end{figure}

\paragraph{Fixed-effects topic proportions (Figure~\ref{fig:topic_fixed}).}
Normalizing $\widehat B$ within each document yields topic proportions
$\widehat\pi_{qn}=\widehat B_{qn}/\sum_{q'}\widehat B_{q'n}$.
Figure~\ref{fig:topic_fixed} displays these proportions chronologically,
highlighting systematic stepwise shifts in topic prevalence across eras.

\paragraph{Inference for era-specific effects (Table~\ref{tab:topic_theta}).}
With no-intercept coding, each coefficient is an era-specific expected latent-score level.
The postwar era is dominated by Topic~3, while Topics~1 and~2 are concentrated in
earlier eras, with close SE--BSE agreement.

\begin{table}[t]
\centering
\caption{Estimated covariate effects for the topic model ($Q=3$).
Format as in Table~\ref{tab:orthodont_theta}.}
\label{tab:topic_theta}
\setlength{\tabcolsep}{4.2pt}
\small
\begin{tabular}{llrrrrr}
\toprule
Period & Topic & Estimate & SE & BSE & $z$ & $p$ \\
\midrule
Early (1775--1865)      & Topic 1 &  46.744 & 12.036 & 11.914 &  3.88 & $<0.001$ \\
Early (1775--1865)      & Topic 2 & 188.558 & 36.392 & 37.162 &  5.18 & $<0.001$ \\
Early (1775--1865)      & Topic 3 &   2.491 &  4.647 &  3.362 &  0.54 & 0.2960 \\
Industrial (1865--1945) & Topic 1 & 164.855 & 24.441 & 24.985 &  6.75 & $<0.001$ \\
Industrial (1865--1945) & Topic 2 &  23.901 & 13.113 & 12.944 &  1.82 & 0.0342 \\
Industrial (1865--1945) & Topic 3 &  35.862 & 14.349 & 14.026 &  2.50 & 0.0062 \\
Postwar (1945--)        & Topic 1 &  16.528 &  9.538 &  8.952 &  1.73 & 0.0416 \\
Postwar (1945--)        & Topic 2 &   6.361 &  7.375 &  5.826 &  0.86 & 0.1942 \\
Postwar (1945--)        & Topic 3 & 186.287 & 11.182 & 10.846 & 16.66 & $<0.001$ \\
\bottomrule
\end{tabular}
\end{table}

\paragraph{Supplementary benchmark: structural topic model (STM) with the same era design.}
As a robustness check, we also fitted a structural topic model (STM) with $Q=3$
\citep{roberts2014,roberts2019} using the same no-intercept era coding (see Appendix~G
for full results).
After aligning topics by representative words, STM corroborates the same qualitative
historical pattern: the civic-rhetoric topic increases sharply from Early to Postwar,
whereas the institutional/legal topic becomes small in the Postwar era.

\paragraph{Summary.}
NMF-RE adds explicit mean-structure inference with non-negative word profiles and
complexity diagnostics to the era signal corroborated by STM (see Appendix~G for
a detailed comparison).
Measurement-side selection is visible in $\widehat X$, which assigns each word to its
dominant topic (institutional/legal, civic-rhetoric, or union/constitutional).
Covariate-side selection through $\Theta$ reveals that each era is characterized
by a distinct dominant topic---Topic~2 (union/constitutional) for the Early era,
Topic~1 (institutional/legal) for the Industrial era, and Topic~3 (civic rhetoric)
for the Postwar era---with nonsignificant entries identifying the absent
era--topic combinations (Table~\ref{tab:topic_theta}).
The random-effects component is effectively inactive ($\mathrm{df}_U/(NQ)$ near zero),
indicating that the temporal pattern is driven by the fixed-effects component.


\section{Discussion}\label{sec:discussion}

This paper proposed NMF-RE, a mean-structure extension of covariate-driven NMF
that combines ridge-regularized random effects with one-step multiplier-bootstrap
inference for $\Theta$, bridging mixed-effects modeling and non-negative low-rank
representations.
The four empirical examples illustrated the method across diverse settings:
longitudinal growth (Orthodont), cognitive test batteries (Holzinger--Swineford),
spatial origin--destination flows (OD--HUB), and presidential address topics.
These applications spanned saturation ratios from nearly zero to over $0.95$,
demonstrating that the dfU-cap and complexity diagnostics remain informative
across contrasting random-effects regimes.

\subsection{Interpretation and practical diagnostics}\label{sec:discussion_diagnostics}
A central practical issue in NMF-RE is the potential for near-saturation when the
random effects are weakly regularized.
The saturation ratio $\mathrm{df}_U/(NQ)$ provides a scale-free summary of the
operating regime, and the dfU-cap prevents penalty collapse and degenerate inference
(Appendix~E demonstrates this failure mode in a controlled stress test).
For transparent reporting, we recommend presenting $\mathrm{df}_U/(NQ)$, the cap ratio,
whether the cap was ever activated, and whether the final solution is binding or
non-binding.

For inference, close agreement between sandwich-based and bootstrap-based standard
errors (SE--BSE agreement) indicates that the one-step linearization is adequate;
all four empirical illustrations exhibited such agreement.
When discrepancies are large, practitioners should check convergence, increase $B$,
examine dfU-cap status, or reduce model complexity.
Near-boundary coefficients should be assessed with directional tests and percentile
summaries rather than two-sided Wald approximations (see Appendix~D for details).

\subsection{Variable selection through non-negative latent structure}\label{sec:discussion_varsel}
As outlined in the Introduction, NMF-RE achieves variable selection
on two complementary sides.
On the \emph{measurement side}, the non-negativity constraint plays a role
analogous to rotation or sparsity penalties in factor analysis, but induces
sparse, parts-based loadings as a structural consequence rather than through an
explicit penalty such as the LASSO.
All four empirical illustrations confirm this mechanism: $\widehat X$
consistently groups observed variables into interpretable components---cognitive
test domains, growth trajectories, geographic regions, or word topics---without a
prespecified loading pattern.

On the \emph{covariate side}, formal inference on $\Theta$ identifies which
covariates significantly affect which components, providing a test-based
covariate selection.
Across the examples, the estimated $\Theta$ reveals component-specific patterns
in which some covariate effects are strongly significant while others are
estimated at exactly zero, yielding a selective structure without an explicit
sparsity penalty.
Together, these two mechanisms allow NMF-RE to serve as a data-driven variable
selection tool in psychometric and related settings, complementing traditional
regularization-based approaches with a constraint-driven, interpretable alternative.

\subsection{Methodological limitations}\label{sec:discussion_limitations}
The proposed inference is explicitly \emph{conditional} on the estimated regularized
solution $(\widehat X,\widehat U)$.
Accordingly, it does not attempt to propagate uncertainty due to estimation of
$X$ and $U$.
This post-regularization perspective is deliberate: it yields stable and scalable
uncertainty quantification for the low-dimensional target $\Theta$ while avoiding
repeated constrained optimization.
At the same time, conditional inference may be conservative or anti-conservative
depending on the extent to which variability in $(\widehat X,\widehat U)$ would
affect the distribution of $\widehat\Theta$ in a given application.
Relatedly, NMF models are only partially identifiable up to scaling and
permutation, and the non-convex objective can have local minima; in practice,
warm starts and multiple initializations may be useful when $Q$ is not small or
when the covariate design is highly imbalanced.

The working Gaussian scale model is used to define scores and information matrices,
while robustness is pursued through sandwich covariance estimation and multiplier
bootstrap resampling of score contributions.
These choices are standard in misspecified likelihood and
post-regularization settings; nevertheless, establishing formal validity under general non-Gaussian data
generating processes and under data-adaptive selection of $(\widehat X,\widehat U)$
remains an important direction for further theoretical work.
Finally, the non-negativity constraint $\Theta\ge 0$ can induce truncation
and skewness for near-boundary coefficients, motivating the boundary-aware
diagnostics discussed in Section~\ref{sec:discussion_diagnostics}.

\subsection{Extensions and future work}\label{sec:discussion_extensions}
Several extensions are natural.
First, developing selection-aware inference that propagates uncertainty from
$(\widehat X,\widehat U)$ into $\Theta$ would clarify when conditional inference
is adequate; however, resampling $X$ is complicated by non-convexity and label
switching.
Second, replacing the squared-loss criterion by count-data likelihoods (e.g., Poisson)
would extend NMF-RE to settings with strongly mean-dependent variance.
Third, boundary-aware interval constructions beyond normal approximations are needed
for principled summaries when many coefficients are near zero.
Fourth, data-driven selection of the rank $Q$ is an open problem shared with all
low-rank models; integrating information-criterion or cross-validation methods into
the NMF-RE framework would reduce reliance on external rank diagnostics.

\subsection{Concluding remarks}\label{sec:discussion_conclusion}
NMF-RE provides a practical bridge between classical mixed-effects ideas and modern
non-negative low-rank representations, supporting a mean-structure--centric strategy
for inference in constraint-based latent-variable models with explicit complexity
control and implicit variable selection.
The empirical illustrations showed that benchmark-specific methods (LMM, factor
analysis, STM) can corroborate NMF-RE findings within their own domains, while
NMF-RE offers a unified framework that accommodates all four settings under a single
model and inference apparatus.

\section*{Statements and Declarations}

\noindent\textbf{Funding.} This work was partly supported by JSPS KAKENHI Grant Numbers 22K11930, 25K15229, 24K03007, 25H00482, the project research fund by the Research Center for Sustainability and Environment at Shiga University, and a research grant from the Fuji Seal Foundation.

\noindent\textbf{Competing Interests.} The author has no relevant financial or non-financial interests to disclose.

\noindent\textbf{Ethical Approval.} Not applicable.

\noindent\textbf{Data Availability.} All empirical data sets used are publicly available through R packages (\texttt{nlme}, \texttt{lavaan}, \texttt{quanteda}) or the Japanese e-Stat portal (\url{https://www.e-stat.go.jp}). All analysis scripts, simulation code, and pre-processed data files needed to reproduce the results in this paper are available at \url{https://github.com/ksatohds/nmfre-paper}.

\noindent\textbf{Code Availability.} The R package \texttt{nmfkc} implementing the covariate-driven NMF used for initialization is publicly available at \url{https://ksatohds.github.io/nmfkc/}.

\noindent\textbf{Use of AI Tools.} Claude Code (Anthropic, Claude Opus 4 model, \url{https://claude.ai/code}, accessed February--March 2026) was used for \LaTeX{} formatting, compilation, cross-reference verification, and manuscript proofreading. No scientific content---including the methodology, mathematical derivations, data analysis, or interpretation of results---was generated by AI tools. The author assumes full responsibility for the accuracy and integrity of all content.

\appendix
\setcounter{secnumdepth}{1}

\section{Practical Calibration of the dfU-Cap}\label{app:calibration}

Let $r^{\max}\equiv \mathrm{df}_U^{\max}/(NQ)$ denote the cap ratio and let
$X_{\mathrm{fix}}$ be the basis obtained from the covariate-driven NMF fit
with $U\equiv 0$ used for initialization.
For a fixed $X$, define the saturation-ratio function
\[
r_X(\lambda)\equiv \frac{\mathrm{df}_U(\lambda)}{NQ}
= \frac{1}{Q}\sum_{q=1}^{Q}\frac{d_q}{d_q+\lambda},
\]
where $d_1,\dots,d_Q$ are the eigenvalues of $X^\top X$.
Computing $r_{X_{\mathrm{fix}}}(\lambda)$ over a log-scale grid yields a simple
``$\lambda\leftrightarrow r$'' lookup before running the full NMF-RE algorithm.
If values $\lambda<\lambda_{\min}$ are deemed implausible for the application,
then setting
\[
\mathrm{df}_U^{\max} = r_{X_{\mathrm{fix}}}(\lambda_{\min})\,NQ
\]
provides a principled cap that avoids trial-and-error scans.
This pre-calibration uses $X_{\mathrm{fix}}$ only to choose
$\mathrm{df}_U^{\max}$; the cap is enforced during optimization using the
current $X$ through $\lambda_{\mathrm{cap}}(X)$.

For transparent reporting, we distinguish (i)~whether the cap was ever activated
during optimization and (ii)~whether the final converged solution is
\emph{binding} ($\mathrm{df}_U=\mathrm{df}_U^{\max}$) or \emph{non-binding}
($\mathrm{df}_U<\mathrm{df}_U^{\max}$).
We report the realized saturation ratio $\mathrm{df}_U/(NQ)$, the cap ratio
$\mathrm{df}_U^{\max}/(NQ)$, the final $\widehat{\lambda}$, and the
activation/binding indicators.

\section{Estimation Algorithm}\label{app:algorithm}

Algorithm~\ref{alg:nmfre} presents the complete block-wise estimation procedure
for NMF-RE with a df-based cap for random effects, as described in
Section~3 of the main text.

\begin{algorithm}[t]
\caption{Block-wise estimation of NMF-RE with a df-based cap for random effects}
\label{alg:nmfre}
\begin{algorithmic}[1]
\Require Non-negative data $Y\in\mathbb{R}_{\ge0}^{P\times N}$, covariates $A\in\mathbb{R}^{K\times N}$,
rank $Q$, initial penalty $\lambda>0$, cap level $\mathrm{df}_U^{\max}$,
tolerance \texttt{tol}, maximum iterations \texttt{maxit}.
\Ensure $\widehat X,\widehat\Theta,\widehat U$ and final $\lambda$.

\Statex \textbf{Initialize:}
\State Fit covariate-driven NMF with $U\equiv0$ to get $(X,\Theta)$; set $U\gets 0$.
\State Normalize $X$ so that $\mathbf{1}_P^\top X=\mathbf{1}_Q^\top$ and absorb scaling into $\Theta$.

\For{$t=1,2,\dots,\texttt{maxit}$}
    \State Compute current objective
    $L_{\text{old}} \gets \|Y - X(\Theta A + U)\|_F^2 + \lambda\|U\|_F^2$.

    \Statex \textbf{(i) Enforce df cap (diagnostic safeguard):}
    \State Compute eigenvalues $d_1,\dots,d_Q$ of $X^\top X$.
    \State Find $\lambda_{\mathrm{cap}}(X)=\inf\{\lambda'>0:\ N\sum_{q=1}^Q \frac{d_q}{d_q+\lambda'}\le \mathrm{df}_U^{\max}\}$
    (e.g., bisection).
    \State $\lambda \gets \max\{\lambda,\lambda_{\mathrm{cap}}(X)\}$.

    \Statex \textbf{(ii) $U$-step (ridge / BLUP-like update):}
    \State $M \gets (X^\top X+\lambda I_Q)^{-1}X^\top$.
    \State $U \gets M\bigl(Y - X(\Theta A)\bigr)$.
    \State Row-center $U$ so that each row has mean zero across units.

    \Statex \textbf{(iii) $X$-step (stabilized multiplicative update + renormalization):}
    \State $B_U \gets \Theta A + U$, \quad $B_U^{+}\gets \max(B_U,0)$ (elementwise).
    \State $X_{\text{cand}} \gets X \odot \bigl(Y(B_U^{+})^\top\bigr)\oslash \bigl(X(B_U^{+}(B_U^{+})^\top)\bigr)$.
    \State Normalize columns: $X_{\text{cand}} \gets X_{\text{cand}}D^{-1}$ so that $\mathbf{1}_P^\top X_{\text{cand}}=\mathbf{1}_Q^\top$.
    \State Rescale $(\Theta,U)\gets (D\Theta,DU)$ to keep $X(\Theta A+U)$ unchanged.

    \Statex \textbf{(iv) $\Theta$-step (stabilized multiplicative update):}
    \State $Y_U \gets Y - X U$, \quad $Y_U^{+}\gets \max(Y_U,0)$ (elementwise).
    \State $\Theta_{\text{cand}} \gets \Theta \odot \bigl(X^\top Y_U^{+}A^\top\bigr)\oslash \bigl((X^\top X\Theta)(AA^\top)\bigr)$.

    \Statex \textbf{(v) Descent safeguard using the true working objective:}
    \State Form candidate $(X,\Theta)\gets (X_{\text{cand}},\Theta_{\text{cand}})$ and compute
    $L_{\text{new}} \gets \|Y - X(\Theta A + U)\|_F^2 + \lambda\|U\|_F^2$.
    \If{$L_{\text{new}} > L_{\text{old}}$}
        \State (Optional) apply damping / rollback, e.g.,
        set $X\gets X_{\text{old}}$, $\Theta\gets \Theta_{\text{old}}$ and skip the update,
        or replace $(X,\Theta)\gets (1-\eta)(X_{\text{old}},\Theta_{\text{old}})+\eta(X_{\text{cand}},\Theta_{\text{cand}})$ with $\eta\in(0,1)$.
    \EndIf

    \Statex \textbf{(vi) Check convergence:}
    \If{$|L_{\text{old}}-L_{\text{new}}|/(L_{\text{old}}+\epsilon) < \texttt{tol}$}
        \State \textbf{break}
    \EndIf
\EndFor

\State \Return $(\widehat X,\widehat\Theta,\widehat U)\gets (X,\Theta,U)$.
\end{algorithmic}
\end{algorithm}

\paragraph{Remark (stabilized multiplicative updates and objective monitoring).}
When $U$ is real-valued, the intermediate matrices $B_U=\Theta A+U$ and $Y_U=Y-XU$ can contain negative entries,
so the classical Euclidean multiplicative updates are not directly applicable.
We therefore use elementwise positive parts $B_U^{+}=\max(B_U,0)$ and $Y_U^{+}=\max(Y_U,0)$ only as stabilized surrogates
to keep multiplicative ratios well-defined and to preserve non-negativity of $X$ and $\Theta$.
Because these stabilized updates are not derived as an exact majorization--minimization scheme for the full objective
$\mathcal{L}(X,\Theta,U)$, we explicitly monitor $\mathcal{L}$ after each block update
and employ a simple descent safeguard (rollback or damping) if an iteration increases the objective.
In our numerical experiments, this safeguard was rarely activated, but it provides a transparent guarantee against pathological steps.

\paragraph{Remark (warm-start variance estimation).}
Algorithm~\ref{alg:nmfre} treats $\lambda>0$ as a given initial penalty,
adjusted only upward by the dfU cap.
In practice, $\lambda=\sigma^2/\tau^2$ depends on the noise variance $\sigma^2$,
which is unknown before fitting.
Our implementation therefore uses a warm-start schedule:
$\sigma^2$ is held at a fixed initial value (e.g., $\sigma^2=1$) for the first
several dozen iterations while the block-wise updates stabilize $(X,\Theta,U)$,
and is then gradually updated toward the current residual mean square
$\|Y-X(\Theta A+U)\|_F^2/(PN)$ via an exponential moving average with a small
learning rate.
Because the dfU cap provides a floor on $\lambda$, this adaptive schedule cannot
drive $\lambda$ to zero; it merely aligns the working variance scale with the data
without destabilizing the optimization.

\section{Asymptotic Justification of the One-Step Map}\label{app:onestep}

Conditional on $(\widehat X,\lambda)$ and with $U$ profiled out by ridge regression,
$\widehat\Theta$ can be viewed as a (possibly misspecified) Z-estimator solving
$\sum_{n=1}^N \mathrm{vec}\{S_n(\Theta)\}=\mathbf{0}$.
Under standard regularity conditions for Z-estimators---smoothness of the score map,
nonsingularity of the Jacobian (equivalently, the reduced information) at the target,
and an initial estimator $\widetilde\Theta$ that is $\sqrt{N}$-consistent and lies in an
$O_p(N^{-1/2})$ neighborhood of the target---a one-step Newton update
\[
\mathrm{vec}(\Theta^{\mathrm{os}})
=
\mathrm{vec}(\widetilde\Theta)
-
\mathcal{I}(\widetilde\Theta)^{-1}
\sum_{n=1}^N \mathrm{vec}\{S_n(\widetilde\Theta)\}
\]
is asymptotically equivalent to the fully iterated Z-estimator (the root of the
estimating equation) in the interior regime, in the sense that
$\sqrt{N}\,(\Theta^{\mathrm{os}}-\widehat\Theta)\to 0$ in probability.
This is a classical consequence of asymptotic linearity for M/Z-estimators and
underlies one-step/debiased constructions in post-regularization inference
\citep[see, e.g.,][]{huber1967,white1980,van2014,javanmard2014,taylor2015}.
In our bootstrap construction, $S^{\ast}(\widehat\Theta)$ in the one-step formula denotes the
multiplier-resampled score (Sections~4.4--4.5 of the main text).
The non-negativity constraint $\Theta\ge0$ is handled by projection of one-step replicates;
near the boundary, the limiting distribution can be non-Gaussian, so directional tests and
percentile-type summaries are more appropriate than symmetric Wald approximations.

\section{Validity, Scope, and Practical Diagnostics}\label{app:validity}

This appendix provides the full discussion of the inferential scope, regularity
conditions, and practical diagnostics summarized in Section~4.7 of the main text.

\paragraph{Scope of inference (conditional target).}
Inference in this paper targets the covariate-effect matrix $\Theta$
\emph{conditional on} the selected regularized representation $(\widehat X,\widehat U)$
and the final ridge level $\lambda$ produced by the estimation algorithm.
Accordingly, sampling variability is assessed through unit-wise (column-wise) score
contributions given fixed covariates $A$, while additional uncertainty due to the
non-convex estimation of $X$ and the regularized fit of $U$ is treated as part of the
post-regularization approximation \citep{white1980,van2014,javanmard2014}.

\paragraph{Estimating-equation viewpoint and robust covariance.}
With $(\widehat X,\lambda)$ treated as fixed and $U$ profiled out by ridge regression,
$\widehat\Theta$ can be viewed as a (possibly misspecified) Z-estimator solving the
estimating equation
\[
\sum_{n=1}^N \mathrm{vec}\{S_n(\Theta)\}=\mathbf{0},
\]
where $S_n(\Theta)$ is the unit-wise score contribution.
Under standard regularity conditions for M/Z-estimators---notably,
(i) independence across units $n$ (or validity of a cluster-robust adaptation),
(ii) finite second moments of the score contributions, and
(iii) nonsingularity of the reduced information matrix---the sandwich covariance estimator
provides large-sample robust standard errors even under a misspecified Gaussian working model
\citep{huber1967,white1980,cameron2008}.

\paragraph{Multiplier (wild) bootstrap as a fast resampling approximation.}
The multiplier (wild) bootstrap resamples the empirical
distribution of the score contributions by random reweighting, avoiding repeated
constrained re-optimization.
Under mild moment conditions, such multiplier schemes consistently approximate the
distribution of smooth functionals of estimating-equation solutions
\citep{mammen1993,cameron2008}.
Combined with the one-step Newton map, this yields a computationally
efficient approximation to resampling-based uncertainty quantification in our
post-regularization setting \citep{van2014,javanmard2014}.

\paragraph{Boundary behavior under $\Theta\ge 0$.}
Because the one-step update is unconstrained, bootstrap replicates are
projected onto the non-negative orthant.
Coefficients near the boundary can therefore exhibit truncation and skewness, so
directional (one-sided) tests and percentile-type bootstrap summaries are often more
appropriate than symmetric Gaussian Wald approximations.
In such boundary regimes, discrepancies between sandwich-based standard errors and
bootstrap-based standard errors can arise mechanically from projection and should not be
interpreted as a failure of the one-step linearization.

\paragraph{Practical diagnostics and reporting.}
For transparent reporting and to guard against near-saturated random-effects updates, we
recommend documenting:
(i) the saturation ratio $\mathrm{df}_U/(NQ)$ and the cap level $\mathrm{df}_U^{\max}/(NQ)$
(and, when relevant, an upper-tail summary such as dfU$_{0.99}$ over iterations or across
restarts);
(ii) whether the dfU-cap was ever activated during optimization and whether the final
solution is binding ($\mathrm{df}_U=\mathrm{df}_U^{\max}$) or non-binding
($\mathrm{df}_U<\mathrm{df}_U^{\max}$), together with the final $\widehat\lambda$;
(iii) agreement between sandwich-based and bootstrap-based standard errors for coefficients
well inside the feasible region; and
(iv) sensitivity to initialization (multiple starts) and convergence behavior when $Q$ is
moderate or large.
These checks help assess whether the selected low-rank representation is sufficiently
stable for the intended conditional inference on $\Theta$.

\section{Stress Test for Random-Effects Saturation}\label{app:stress}

This appendix reports an additional targeted stress experiment designed to verify
that the dfU-cap functions as a safety device against random-effects saturation.
The stress test uses a more flexible setting ($P=4$, $Q=3$, $N=100$) than the
baseline simulation in the main text, where the working random-effects penalty
is intentionally weakened.
\emph{This experiment is designed as a failure-mode demonstration (near-saturation and $\widehat\lambda$ collapse), not as a benchmark of inferential accuracy under well-regularized fits.}
Specifically, we initialize the variance ratio so that $\lambda=\sigma^2/\tau^2$ is extremely small by using
$\sigma^2_{\text{fit}}=1$ and $\tau^2_{\text{fit}}=1000$, which (without safeguards) promotes a near-saturated
$U$ update and can drive $\widehat\lambda$ toward zero.

For this stress test we use a synthetic three-trend basis matrix $X_{\text{true}}\in\mathbb{R}^{4\times 3}$ with
column sums normalized to one:
\[
X_{\text{true}}=
\bigl(
(0.45,0.30,0.15,0.10)^\top,\;
(0.25,0.25,0.25,0.25)^\top,\;
(0.10,0.15,0.30,0.45)^\top
\bigr),
\qquad
\mathbf{1}_P^\top X_{\text{true}}=\mathbf{1}_3^\top .
\]
The intercept is split evenly across the three trends, and the male effect is placed on Trend1 only:
$\Theta_{\text{true}}[\text{Trend}1,\texttt{male}]=9.4285$ and
$\Theta_{\text{true}}[\text{Trend}2,\texttt{male}]=\Theta_{\text{true}}[\text{Trend}3,\texttt{male}]=0$, while
$\Theta_{\text{true}}[\text{Trend}q,\texttt{intercept}]=90.502/3$ for $q=1,2,3$.
The data are generated from the same data-generating process as the baseline simulation with $U\in\mathbb{R}^{3\times N}$.

We compare three dfU-cap settings: a strict cap $\mathrm{df}_U^{\max}/(NQ)=0.21$, a loose cap $0.99$, and
no cap.
The goal is to demonstrate what happens when $U$ becomes (nearly) saturated and whether dfU-cap prevents
the resulting degeneracy in inference for $\Theta$.

\begin{table}[t]
\centering
\caption{Stress test for random-effects saturation ($P=4$, $Q=3$, $N=100$; $R=1000$; $B=1000$).
The fitting routine is initialized with an intentionally weak random-effects penalty
($\sigma^2_{\text{fit}}=1$, $\tau^2_{\text{fit}}=1000$), which (without safeguards) encourages $\widehat\lambda$
to approach zero and $U$ to become nearly saturated.
Hypothesis indicates the true-parameter regime:
Null ($\theta_{\texttt{male}}=0$) or Alternative ($\theta_{\texttt{male}}\neq 0$).
dfU$_{0.99}$ is the 99th percentile (over Monte Carlo runs) of the saturation ratio $\mathrm{df}_U/(NQ)$, and
MeanLambda is the Monte Carlo mean of the fitted penalty $\widehat\lambda$.
Reject is the BSE-based Wald rejection rate (one-sided for Null, two-sided for Alternative);
Cover is the nominal 95\% Wald coverage.
\emph{This table is intended to demonstrate the degeneracy induced by near-saturation and the protective role of dfU-cap, rather than to provide a general performance comparison across caps.}}
\label{tab:sim_dfUcap_stressQ3}

\setlength{\tabcolsep}{3.2pt}
\small
\begin{tabular}{lllrrrrrr}
\toprule
Hypothesis & $r^{\max}$ & Error &
Bias & SD &
dfU$_{0.99}$ & MeanLambda &
Reject & Cover \\
\midrule
Null        & 0.21 & Gaussian     & 0.131 & 0.220 & 0.210 & 1.5988    & 0.050 & 0.970 \\
Null        & 0.99 & Gaussian     & 0.093 & 0.106 & 0.990 & $<$0.001  & 0.000 & 1.000 \\
Null        & off  & Gaussian     & 0.095 & 0.115 & 0.999 & $<$0.001  & 0.000 & 1.000 \\
Null        & 0.21 & Exp-centered & 0.105 & 0.209 & 0.210 & 1.6462    & 0.020 & 0.990 \\
Null        & 0.99 & Exp-centered & 0.089 & 0.102 & 0.990 & $<$0.001  & 0.000 & 1.000 \\
Null        & off  & Exp-centered & 0.090 & 0.104 & 0.999 & $<$0.001  & 0.000 & 1.000 \\
\midrule
Alternative & 0.21 & Gaussian     &-2.477 & 1.002 & 0.210 & 1.3792    & 0.960 & 0.250 \\
Alternative & 0.99 & Gaussian     &-1.151 & 0.998 & 0.990 & $<$0.001  & 1.000 & 0.650 \\
Alternative & off  & Gaussian     &-1.710 & 0.785 & 0.999 & $<$0.001  & 1.000 & 0.110 \\
Alternative & 0.21 & Exp-centered &-2.388 & 1.052 & 0.210 & 1.3345    & 0.980 & 0.280 \\
Alternative & 0.99 & Exp-centered &-1.118 & 0.797 & 0.990 & $<$0.001  & 1.000 & 0.640 \\
Alternative & off  & Exp-centered &-1.722 & 0.739 & 0.999 & $<$0.001  & 1.000 & 0.100 \\
\bottomrule
\end{tabular}
\end{table}

\paragraph{Failure mode without dfU-cap.}
Table~\ref{tab:sim_dfUcap_stressQ3} is a targeted failure-mode demonstration of random-effects saturation.
With the strict cap ($0.21$), dfU$_{0.99}$ stays near $0.21$ and the fitted penalty remains in a stable regime
(MeanLambda around $1.3$--$1.7$), keeping the $U$-update away from near-saturation.
In this regime, the BSE-based one-sided size under the Null remains close to 5\% and Wald coverage is near
(or slightly above) nominal.

When the cap is loose ($0.99$) or turned off, dfU$_{0.99}$ approaches one and the fitted penalty collapses
(MeanLambda below $10^{-3}$), corresponding to an almost unpenalized $U$ update.
Under the Null this produces \emph{degenerate} inference: the Wald test essentially never rejects and coverage
inflates to 1.00, indicating that $U$ absorbs nearly all idiosyncratic variation and leaves little stable
information for $\Theta$.

Under the Alternative, rejection remains high across caps, but coverage is poor
in all three settings.
Notably, coverage under the strict cap ($0.250$/$0.280$) is \emph{lower} than under
the loose cap ($0.650$/$0.640$), because the strict cap constrains $U$ more heavily
and thereby redirects additional bias into $\widehat\Theta$.
This is expected in the present stress design: the intentionally weak initial penalty
combined with a higher-dimensional random-effects space ($Q=3$) creates a regime
in which $\widehat\theta$ is substantially biased under all cap settings, so
coverage alone is not the appropriate success criterion here.
What the table \emph{does} demonstrate is the contrast between the strict cap and
the loose/off settings: without the cap, the fitted penalty collapses
($\widehat\lambda\approx 0$) and inference becomes degenerate, whereas the strict
cap keeps $\widehat\lambda$ in a stable regime and preserves meaningful (albeit
imperfect) uncertainty quantification.
Under well-calibrated conditions---as in the baseline $Q=1$ simulation
(Table~\ref{tab:sim_male})---both size and coverage are near nominal.
Overall, the stress test shows that dfU-cap is a practically important safeguard:
it prevents $\widehat\lambda$ collapse and blocks pathological inference in regimes
that would otherwise admit near-saturated $U$.

\paragraph{Random-effects complexity monitoring.}
In the baseline $Q=1$ design reported in Table~\ref{tab:sim_male}, the mean saturation ratio
$\mathrm{df}_U/(NQ)$ stays near $0.20$ under the default cap $\mathrm{df}_U^{\max}/(NQ)=0.21$,
indicating that $U$ is kept in a moderately regularized regime.
The $Q=3$ stress experiment in Table~\ref{tab:sim_dfUcap_stressQ3} shows why this monitoring matters:
if $\mathrm{df}_U/(NQ)$ is allowed to approach one (loose/off cap), the fitted penalty can collapse and inference
for $\Theta$ can become degenerate or severely distorted.

\section{Detailed Simulation Results Discussion}\label{app:sim_discussion}

This appendix presents the full Monte Carlo results and provides a detailed
interpretation.

\begin{table}[t]
\centering
\caption{Monte Carlo results for the male effect $\theta_{\texttt{male}}$ under the
Orthodont-based design ($P=4$, $Q=1$), with $R=1000$ replicates and multiplier bootstrap
size $B=1000$.
Hypothesis indicates the true-parameter regime:
Null ($\theta_{\texttt{male}}=0$) or Alternative ($\theta_{\texttt{male}}\neq 0$).
Reject is the Wald rejection rate (one-sided for Null, two-sided for Alternative).
Mean is the Monte Carlo average of the standard error estimate (SE or BSE).
Entries of the form ``a/b'' report ``SE-based / BSE-based'' quantities.
PctCover is the coverage of the 95\% percentile bootstrap CI, and dfU is the Monte
Carlo mean of $\mathrm{df}_U/(NQ)$.}
\label{tab:sim_male}

\setlength{\tabcolsep}{3.2pt}
\small
\begin{tabular}{llrrrrrrrr}
\toprule
$N$ & Hypothesis &
Bias & SD & RMSE &
Mean & Reject & Cover & PctCover & dfU \\
\midrule
\multicolumn{10}{l}{\emph{Gaussian errors}}\\
27  & Null        & 0.373 & 0.526 & 0.645 & 0.853/0.572 & 0.064/0.082 & 0.956/0.951 & 0.956 & 0.203 \\
27  & Alternative &-0.013 & 0.919 & 0.919 & 0.842/0.826 & 1.000/1.000 & 0.912/0.902 & 0.910 & 0.201 \\
100 & Null        & 0.179 & 0.246 & 0.304 & 0.443/0.304 & 0.048/0.055 & 0.977/0.970 & 0.978 & 0.204 \\
100 & Alternative &-0.010 & 0.430 & 0.430 & 0.442/0.439 & 1.000/1.000 & 0.953/0.952 & 0.952 & 0.201 \\
200 & Null        & 0.144 & 0.185 & 0.235 & 0.324/0.226 & 0.057/0.062 & 0.970/0.966 & 0.970 & 0.206 \\
200 & Alternative &-0.011 & 0.325 & 0.325 & 0.324/0.323 & 1.000/1.000 & 0.949/0.948 & 0.946 & 0.201 \\
\multicolumn{10}{l}{\emph{Centered exponential errors}}\\
27  & Null        & 0.333 & 0.473 & 0.579 & 0.860/0.567 & 0.057/0.076 & 0.969/0.963 & 0.967 & 0.201 \\
27  & Alternative &-0.025 & 0.854 & 0.854 & 0.850/0.835 & 1.000/1.000 & 0.931/0.926 & 0.928 & 0.201 \\
100 & Null        & 0.186 & 0.267 & 0.326 & 0.444/0.304 & 0.063/0.074 & 0.967/0.963 & 0.966 & 0.202 \\
100 & Alternative &-0.011 & 0.457 & 0.457 & 0.443/0.441 & 1.000/1.000 & 0.939/0.937 & 0.938 & 0.201 \\
200 & Null        & 0.155 & 0.191 & 0.246 & 0.323/0.227 & 0.059/0.071 & 0.966/0.961 & 0.964 & 0.204 \\
200 & Alternative & 0.009 & 0.328 & 0.328 & 0.323/0.322 & 1.000/1.000 & 0.940/0.941 & 0.938 & 0.201 \\
\bottomrule
\end{tabular}
\end{table}

\paragraph{What SE--BSE agreement does (and does not) validate.}
When coefficients lie in the interior of the feasible region, agreement between
SE and BSE provides direct evidence that the estimator is well captured by its
local linearization, which is precisely the approximation used by the one-step
multiplier bootstrap.
However, in boundary regimes induced by $\Theta\ge0$, SE--BSE agreement is not
the appropriate success criterion: projection of one-step replicates to
$\Theta\ge0$ alters the finite-sample distribution by adding truncation at zero.
In that regime, the relevant validation is whether directional testing and
interval summaries retain reasonable size and coverage.

\paragraph{Interior regime (Alternative): local linearization works well.}
Under the Alternative, Mean(SE) and Mean(BSE) are close for both error distributions and
all $N$, and coverages approach nominal levels as $N$ increases.
This behavior is consistent with the intended post-regularization
interpretation: once the selected low-rank structure is stable and coefficients
are away from the boundary, the one-step bootstrap provides an accurate local
uncertainty approximation.

\paragraph{Boundary regime (Null): projection induces systematic SE--BSE differences.}
Under the Null, Mean(BSE) is systematically smaller than Mean(SE).
This is expected: the one-step bootstrap replicates are projected onto
$\Theta\ge0$, which concentrates mass at zero and reduces the empirical spread
compared to a symmetric Gaussian Wald approximation.
Despite this boundary-induced truncation and the resulting positive bias in
$\widehat\theta_{\texttt{male}}$, the one-sided rejection rates remain close to
the nominal 5\% level and Wald-type coverages are slightly conservative across
$N$.
Percentile bootstrap intervals show similar coverages, indicating that the
projected one-step bootstrap remains practically usable as a boundary-aware
uncertainty summary in this setting.

\paragraph{Robustness to skewed observation noise.}
Replacing Gaussian errors by centered exponential errors yields qualitatively
similar bias, SD, RMSE, and coverage patterns.
Under the Null, the BSE-based one-sided test becomes mildly more liberal in finite
samples, but the deviation is modest and does not worsen with increasing $N$.
Overall, this suggests that the proposed post-regularization inference retains
useful stability under moderate deviations from Gaussianity of the measurement
noise, at least in this Orthodont-like low-rank design.

\section{Supplementary Benchmark Comparisons}\label{app:benchmarks}
\renewcommand{\thesubsection}{\theappendix.\arabic{subsection}}

This appendix collects benchmark comparisons with established methods for each of the
four empirical illustrations in the main text.
These comparisons contextualize the NMF-RE results relative to standard approaches
and are not intended as head-to-head performance evaluations.

\subsection{Orthodont: Linear mixed-effects modeling (LMM)}\label{app:lmm}

A standard linear mixed-effects model (LMM) for the Orthodont data specifies a
\emph{prescribed} time basis (typically an intercept and a linear age term) and
then introduces random effects on that basis.
For subject $n$ at age $t\in\{8,10,12,14\}$, a common specification is
\[
y_{nt}
=
(\beta_0+\beta_{\mathrm{sex}}\mathrm{male}_n)
+
(\beta_1+\beta_{\mathrm{age:sex}}\mathrm{male}_n)\,t
+
(b_{0n}+b_{1n}t)
+\varepsilon_{nt},
\]
so the analyst must choose and justify a time design \emph{a priori} (and add
curvature terms such as $t^2$ or splines if needed).
With only four observation times, this design choice can be consequential.

To make the sex main effect interpretable within the observed age range, we also
fit the centered-age form with $t_c=t-11$:
\[
y_{nt}
=
\beta_0+\beta_1 t_c+\beta_2\mathrm{male}_n+\beta_3 t_c\,\mathrm{male}_n
+(b_{0n}+b_{1n}t_c)+\varepsilon_{nt}.
\]
In this parameterization, $\beta_2$ represents the male effect on distance at age 11.
The fitted model indicates that males have a higher mean distance
by about 2.32\,mm ($\hat\beta_2=2.321$, $p=0.0054$), and that the growth rate is
also larger for males through the age-by-sex interaction
($\hat\beta_3=0.305$, $p=0.0264$), yielding slopes of 0.784 (males) and
0.480 (females) mm/year.
Random-effect estimates suggest non-negligible between-subject heterogeneity in
both level and slope (SDs about 1.83 and 0.18, respectively) with residual SD
about 1.31.
The LMM here serves as an orientation benchmark confirming that the Orthodont data exhibit
both substantial subject-level heterogeneity and sex-dependent mean structure.

\subsection{Holzinger--Swineford: Confirmatory MIMIC / latent regression in SEM}\label{app:mimic}

To benchmark the NMF-RE mean-structure approach against a classical covariance-based
SEM formulation, we fitted a confirmatory MIMIC (multiple indicators multiple causes)
model \citep{joreskoggoldberger1975,bollen1989} using the \texttt{lavaan} package
\citep{lavaan}.
Following the textbook three-factor specification, the nine tests were assigned to
\texttt{visual} ($x_1$--$x_3$), \texttt{textual} ($x_4$--$x_6$), and
\texttt{speed} ($x_7$--$x_9$), and these three factors were regressed on the covariates.
We used robust (sandwich) standard errors (MLR) and fixed factor variances to one
(\texttt{std.lv=TRUE}) to stabilize interpretation of the regression coefficients.

Table~\ref{tab:hs1939_mimic} reports the latent regression coefficients.
In this confirmatory analysis, $\texttt{sex.2}$ is significantly associated with the
\texttt{visual} factor, while $\texttt{school.GW}$ is strongly associated with the
\texttt{textual} factor; $\texttt{age.rev}$ shows factor-specific effects, including a
clear association with \texttt{speed}.
Joint Wald tests that constrain each covariate's three regression coefficients to zero
simultaneously reject for $\texttt{age.rev}$, $\texttt{sex.2}$, and $\texttt{school.GW}$,
indicating that all three covariates have nontrivial impact on the latent structure under
the confirmatory specification.
As an overall reference, the fitted MIMIC model yields CFI=0.893, TLI=0.840,
RMSEA=0.092, and SRMR=0.061.

\begin{table}[t]
\centering
\caption{Confirmatory MIMIC / latent regression results (SEM) for the
Holzinger--Swineford data: regression of three textbook factors on covariates.
Standard errors are robust (sandwich) under MLR. P-values are two-sided.}
\label{tab:hs1939_mimic}
\setlength{\tabcolsep}{4.2pt}
\small
\begin{tabular}{llrrrr}
\toprule
Covariate & Factor & Estimate & SE & $z$ & $p$ \\
\midrule
age.rev    & visual  &  0.075 & 0.075 &  0.998 & 0.318 \\
age.rev    & textual &  0.183 & 0.059 &  3.131 & 0.002 \\
age.rev    & speed   & -0.264 & 0.073 & -3.605 & $<0.001$ \\
sex.2      & visual  & -0.427 & 0.162 & -2.642 & 0.008 \\
sex.2      & textual &  0.119 & 0.128 &  0.934 & 0.350 \\
sex.2      & speed   &  0.166 & 0.146 &  1.138 & 0.255 \\
school.GW  & visual  & -0.259 & 0.184 & -1.404 & 0.160 \\
school.GW  & textual &  0.521 & 0.129 &  4.053 & $<0.001$ \\
school.GW  & speed   & -0.139 & 0.151 & -0.922 & 0.357 \\
\bottomrule
\end{tabular}
\end{table}

\subsection{OD--HUB: Projection pursuit regression (PPR)}\label{app:ppr}

As a flexible regression benchmark that does not impose a low-rank factorization, we also fitted
multivariate projection pursuit regression (PPR) \citep{friedman1981,friedman1984}.
Treating destinations as observations, we regress the 43-dimensional response
$\boldsymbol y_n\in\mathbb{R}^{43}$ on the 4-dimensional hub-flow predictor
$\boldsymbol a_n\in\mathbb{R}^4$ using $M=4$ ridge terms:
\[
\boldsymbol y_n
=
\sum_{m=1}^M \boldsymbol\beta_m\, g_m(\boldsymbol\alpha_m^\top \boldsymbol a_n)
+\boldsymbol\varepsilon_n,
\]
where $\boldsymbol\alpha_m\in\mathbb{R}^4$ are projection directions shared across responses,
$g_m(\cdot)$ are smooth univariate ridge functions, and $\boldsymbol\beta_m\in\mathbb{R}^{43}$
are response-specific weights.
Using \texttt{ppr} in \textsf{R} with $M=4$ terms (allowing up to 10 candidate terms internally),
the fitted model achieved a moderate multivariate fit on the training data
(overall $R^2=0.632$ when aggregating squared errors over all origins and destinations).
The estimated projection directions typically mix the four hubs within each term (e.g., one term
loads heavily on Osaka and Fukuoka while another mixes Tokyo and Aichi), so interpretability hinges
on ridge-function diagnostics rather than a direct hub-to-component correspondence.

\subsection{Topic model: Structural topic model (STM)}\label{app:stm}

As a robustness check using an off-the-shelf probabilistic topic model, we also fitted a
structural topic model (STM) with three topics using the \texttt{stm} package
\citep{roberts2014,roberts2019}.
To mirror the NMF-RE covariate coding, we used the prevalence formula
\[
\texttt{prevalence} \;=\; \texttt{\textasciitilde\ 0 + era},
\]
so that STM estimates an era-specific expected topic prevalence level for each era
(Early, Industrial, Postwar) \emph{without an intercept}.
Table~\ref{tab:topic_stm_level} reports these era-level prevalence estimates with
simulation-based standard errors under global uncertainty; p-values are two-sided.
Under this parameterization, small (near-zero) coefficients indicate that a topic is
nearly absent in an era, which is directly comparable (qualitatively) to the
``near-zero'' interpretation under the NMF-RE mean-structure coding.

Because topic labels are identifiable only up to permutation, we aligned STM topics to the
NMF-RE topic labels by inspecting representative words (e.g., via \texttt{labelTopics} and
$\Pr(\mathrm{topic}\mid\mathrm{word})$).
In this corpus, STM Topic~2 corresponds to our Topic~1 (institutional/legal;
\emph{constitution/laws/congress}), STM Topic~3 aligns with our Topic~2
(union/constitutional/foreign; \emph{war/union/foreign}), and STM Topic~1 corresponds to our Topic~3
(modern civic rhetoric; \emph{today/america/freedom/world}).
Accordingly, Table~\ref{tab:topic_stm_level} reports STM results after relabeling topics to match
the NMF-RE Topic~1--3 order.

With this alignment, STM corroborates the same qualitative historical pattern.
The civic-rhetoric topic (Topic~3) increases sharply from Early to Postwar, whereas the
institutional/legal topic (Topic~1) becomes small in the Postwar era.
The union/foreign topic (Topic~2) declines after the Early era but remains non-negligible.

\begin{table}[t]
\centering
\caption{STM era-level topic prevalence estimates for the inaugural-address corpus ($K=3$),
fitted with \texttt{prevalence} $=\texttt{\textasciitilde\ 0 + era}$ (no intercept).
Topics are relabeled to align with the NMF-RE Topic~1--3 order (STM 2$\to$Topic~1,
STM 3$\to$Topic~2, STM 1$\to$Topic~3).
Standard errors are simulation-based under global uncertainty; p-values are two-sided.}
\label{tab:topic_stm_level}
\setlength{\tabcolsep}{4.2pt}
\small
\begin{tabular}{llrrrr}
\toprule
Covariate & Topic & Estimate & SE & $t$ & $p$ \\
\midrule
Early (1775--1865)      & Topic 1 & 0.51913 & 0.05369 &  9.669 & $<0.001$ \\
Early (1775--1865)      & Topic 2 & 0.40326 & 0.04381 &  9.205 & $<0.001$ \\
Early (1775--1865)      & Topic 3 & 0.07741 & 0.03688 &  2.099 & 0.040 \\
Industrial (1865--1945) & Topic 1 & 0.47425 & 0.05336 &  8.887 & $<0.001$ \\
Industrial (1865--1945) & Topic 2 & 0.19953 & 0.04014 &  4.971 & $<0.001$ \\
Industrial (1865--1945) & Topic 3 & 0.32591 & 0.04181 &  7.796 & $<0.001$ \\
Postwar (1945--)        & Topic 1 & 0.05729 & 0.04878 &  1.175 & 0.245 \\
Postwar (1945--)        & Topic 2 & 0.15039 & 0.04002 &  3.758 & $<0.001$ \\
Postwar (1945--)        & Topic 3 & 0.79233 & 0.04005 & 19.782 & $<0.001$ \\
\bottomrule
\end{tabular}
\end{table}

\clearpage
\begin{center}\textbf{References}\end{center}
\printbibliography[heading=none]

\end{document}